\documentclass[aps,prl,twocolumn,superscriptaddress,notitlepage]{revtex4-1}
\usepackage{graphicx}
\usepackage{upgreek}
\usepackage[linktoc=page,colorlinks,urlcolor=blue,citecolor=blue,linkcolor=blue]{hyperref}
\usepackage{amssymb,amsmath}
\usepackage{graphicx}
\usepackage{natbib}
\usepackage{mathrsfs}
\usepackage{overpic}
\usepackage{wrapfig}
\usepackage{xcolor}
\definecolor{darkblue}{rgb}{0.0 0.0 0.78}
\definecolor{darkred}{rgb}{0.5 0.0 0.0}

\usepackage{gensymb}

\usepackage{ulem}

\usepackage{pdfpages}

\makeatletter
\AtBeginDocument{\let\LS@rot\@undefined}
\makeatother

\providecommand{\e}[1]{\ensuremath{\times 10^{#1}}}

\usepackage{color}
\usepackage[letterpaper,textwidth=7in,top=.75in,bottom=.75in]{geometry}

\newcommand{\huPhys}{Department of Physics, Harvard University, Cambridge, MA 02138, USA}
\newcommand{\CfA}{Harvard-Smithsonian Center for Astrophysics, Cambridge, MA 02138, USA}
\newcommand{\cbs}{Center for Brain Science, Harvard University, Cambridge, MA 02138, USA}
\newcommand{\ChiPhys}{Department of Physics, University of Chicago, Chicago, IL 60637, USA}
\newcommand{\EPS}{Department of Earth and Planetary Sciences, Harvard University, Cambridge, MA 02138, USA}

\begin{document}

\title{Imaging crystal stress in diamond using ensembles of nitrogen-vacancy centers}
\date{\today}

\author{P. Kehayias} \altaffiliation{Current address: Sandia National Laboratories, Albuquerque, NM 87123, USA} \affiliation{\huPhys} \affiliation{\CfA} 
\author{M. J. Turner} \affiliation{\huPhys} \affiliation{\cbs}
\author{R. Trubko}  \affiliation{\huPhys}\affiliation{\EPS}
\author{J. M. Schloss} \affiliation{\cbs} \affiliation{Department of Physics, Massachusetts Institute of Technology, Cambridge, MA 02139, USA}
\author{C. A. Hart}  \affiliation{\huPhys}
\author{M. Wesson}  \affiliation{\ChiPhys}
\author{D. R. Glenn}  \affiliation{\huPhys}
\author{R. L. Walsworth}  \affiliation{\huPhys} \affiliation{\CfA} \affiliation{\cbs}

\begin{abstract}
We present a micrometer-resolution millimeter-field-of-view stress imaging method for diamonds containing a thin surface layer of nitrogen vacancy (NV) centers. In this method, we reconstruct stress tensor elements over a two-dimensional field of view from NV optically-detected magnetic resonance (ODMR) spectra. We use this technique to study how stress inhomogeneity affects NV magnetometry performance, and show how NV stress imaging is a useful and direct way to assess these effects. This new tool for mapping stress in diamond will aid optimization of NV-diamond sensing, with wide-ranging applications in the physical and life sciences. 
\end{abstract}

\maketitle

Nitrogen-vacancy (NV) color centers in diamond are an increasingly popular tool for sensing and imaging electromagnetic fields and temperature, with wide-ranging applications. In particular, widefield 2D magnetic imaging using ensembles of NV centers can provide micrometer spatial resolution and millimeter field-of-view in ambient conditions, enabling investigations of condensed-matter physics, paleomagnetism, and biomagnetism problems \cite{heziVortices, tetienneGraphene, QDM1ggg, nv_bacteria, hemozoin}. However, one limitation to an NV magnetic imager's sensitivity is intrinsic diamond stress variation, which inhomogeneously shifts the NV ground-state resonance frequencies and spoils the NV spin dephasing time  \cite{P1DQ}. Diamond crystal stress and strain are therefore important to understand and minimize when optimizing NV magnetometry \cite{barryNeurons, kasper_IR} and magnetic microscopy.

\begin{figure}[th]
\begin{center}
\includegraphics[width=\columnwidth]{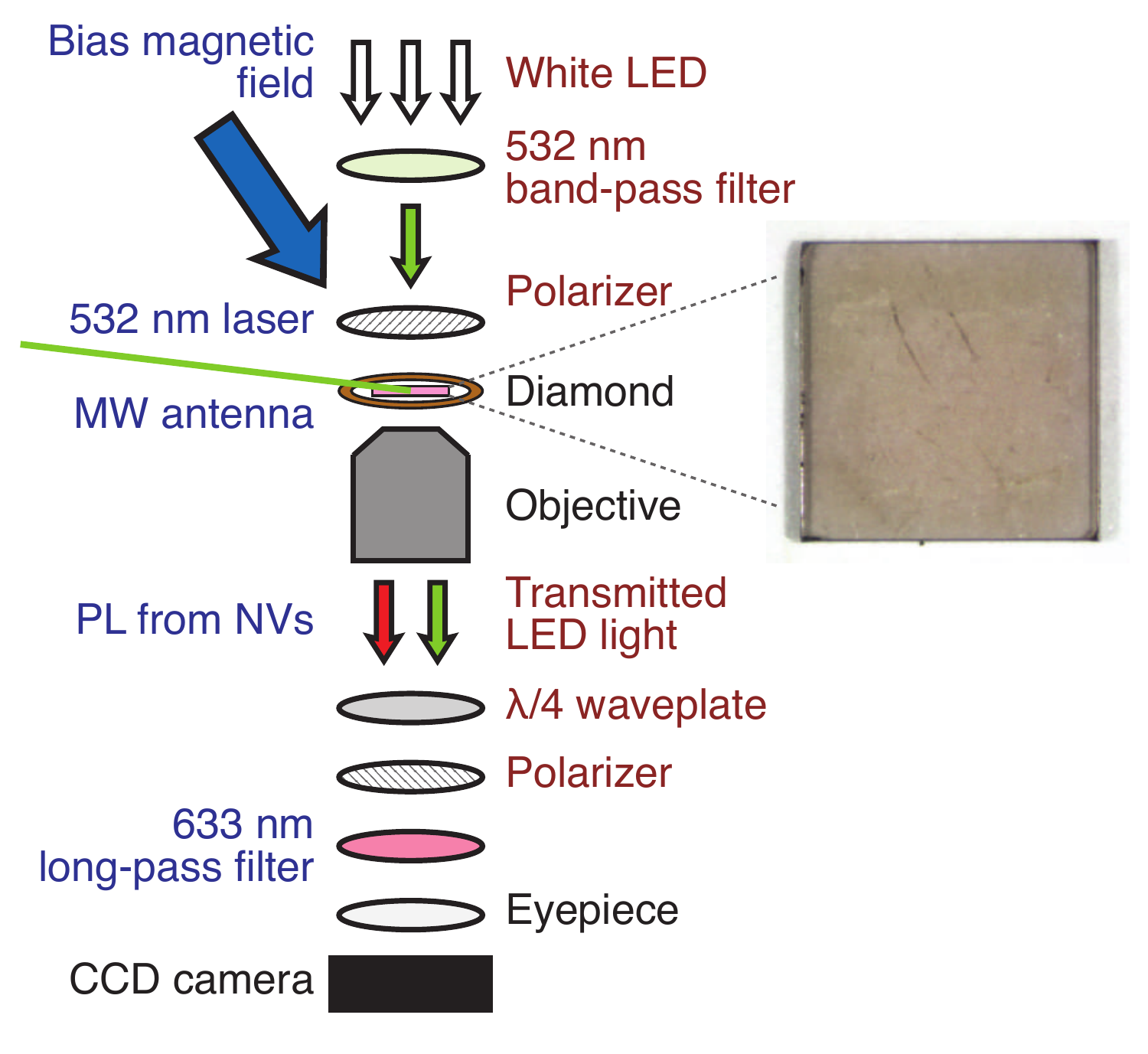}
\end{center}
\caption{\label{setupFig}
Schematic of the combined NV stress  and birefringence imager. The NV stress imager (\textcolor{darkblue}{blue} labels) uses a 532 nm laser to illuminate the diamond, an applied microwave field to drive transitions between NV ground-state sublevels, and a bias magnetic field. The birefringence imager (\textcolor{darkred}{maroon} labels) uses an LED illuminator, two linear polarizers, and a quarter-wave plate. Both imagers use the same microscope and CCD camera (black labels) to collect and image the transmitted light. The photograph on the right shows the diamond Sample A studied in this work.
}
\end{figure}

In this work we use an ensemble NV surface layer to image diamond stress across a millimeter-scale field of view, and explore how stress inhomogeneity impedes NV magnetic microscopy. A 532 nm laser illuminates the micrometer-scale NV layer at the top surface of a diamond chip (4 mm $\times$ 4 mm wide and 0.5 mm thick) and an optical microscope images the spin-state-dependent NV fluorescence onto a camera (Fig.~\ref{setupFig}). Probing the transition frequencies between NV ground-state sublevels by sweeping the frequency of an applied microwave field yields an optically-detected magnetic resonance (ODMR) spectrum in each pixel.  From the resulting 2D map of NV resonance frequencies, we extract magnetic field components and crystal stress tensor elements (which have units of pressure). As crystal stress and strain are related through Young's modulus (1050-1210 GPa for diamond \cite{synthDiamondBook}), we refer to the crystal defects that induce stress within the diamond (shifting the NV ground-state sublevels and causing birefringence) as strain defects. We first demonstrate the NV stress imaging technique with diamond Sample A, which contains a nitrogen-rich layer (25 ppm, 13 $\upmu$m thick) grown on an electronic-grade single-crystal substrate with ppb nitrogen density. This sample was electron-irradiated and annealed to increase the NV density. We also apply the NV stress imaging technique to several other diamonds (Samples B through J), which also exhibit a variety of strain defects (see Supplementary Material \cite{suppl}).

Previous diamond strain imaging studies used X-ray topography, Raman spectroscopy, cathodoluminescence, and birefringence to characterize diamond strain and how it affects diamond applications  \cite{FRIEL2009808, crisci2011, hanleyCath}. By comparison, NV stress imaging gives a more direct characterization of how diamond stress inhomogeneity affects NV magnetic imaging, as both techniques probe the NV ODMR frequencies. In addition, NV stress imaging yields quantitative maps of the diamond stress tensor components localized in the NV layer with micrometer resolution \cite{hollenbergStrain}. The stress tensor reconstruction can help identify how strain features formed during diamond sample preparation and thereby inform future sample fabrication. Finally, high-resolution NV stress imaging is essential in ongoing efforts to identify damage tracks from recoiling carbon nuclei to search for high-energy particle collisions in diamond \cite{nvWIMPdetector}.

In the following sections, we describe NV stress imaging and compare NV-based and birefringence images acquired with the same optical microscope. We next consider how stress inhomogeneity compromises NV magnetometer sensitivity, and then present a survey of common strain defects found in fabricated diamond and their impacts on magnetic microscopy.

\section{Widefield NV Stress imaging}
\begin{figure}[htbp]
\begin{center}
\begin{overpic}[width=\columnwidth]{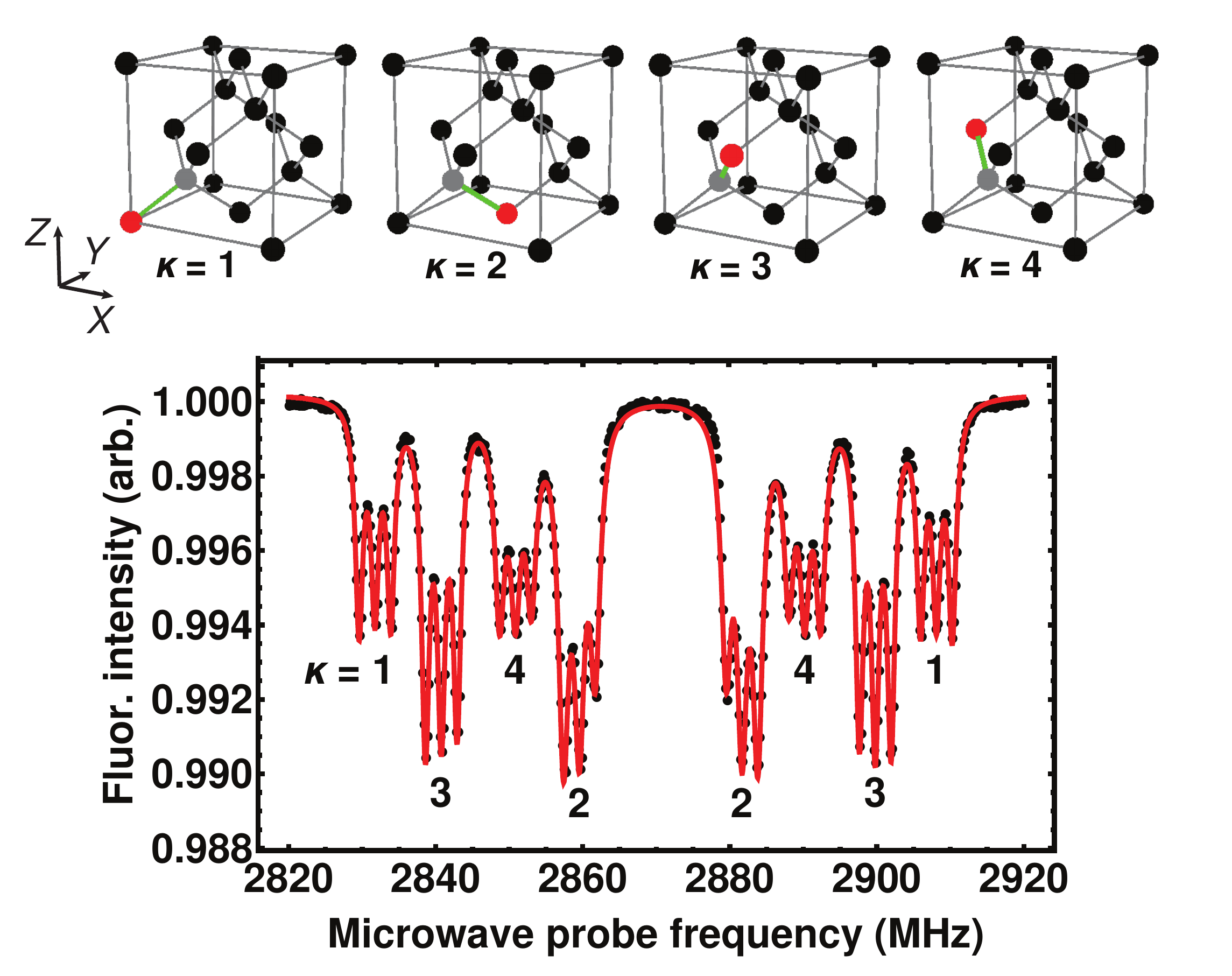}
\put(2,76){\textsf{\Large a}}
\put(2,47){\textsf{\Large b}}
\end{overpic}
\end{center}
\caption{\label{axesFig}
(a) NV centers in the diamond lattice, with the four N-V axes shown in green \cite{maletinskyStrainTerms}. Carbon atoms are black, nitrogen atoms are red, and vacancies are gray.
(b) Example ODMR spectrum with $\vec{B} = \{220, 593, 1520\}$ $\upmu$T in the diamond chip coordinate system (fit function plotted in red). The labels indicate the resonances from the different NV orientations. Each NV resonance is split into three lines due to hyperfine interactions with the spin-1 $^{14}$N nucleus.
}
\end{figure}

The NV center in diamond consists  of a substitutional nitrogen atom in the carbon lattice adjacent to a vacancy (Fig.~\ref{axesFig}a). It has an electronic spin-triplet ground state ($S=1$) with magnetic sublevels $m_s = \{-1, 0, +1 \}$. The sublevel energies shift in response to local magnetic fields, crystal stress, temperature changes, and electric fields. We measure these energy (i.e., frequency) shifts using ODMR spectroscopy, where a resonant microwave field induces transitions between the $m_s = 0$ and $\pm$1 sublevels and causes reduced NV fluorescence under continuous illumination by 532 nm laser light (Fig.~\ref{axesFig}b). Each NV is oriented along one of four crystallographic directions (labeled with the index $\kappa = \{1,2,3,4\}$). An NV ensemble usually contains an equal number of NVs for each $\kappa$. The ODMR spectra from all NV orientations yields the information to reconstruct stress tensor elements and vector magnetic field components \cite{QDM1ggg}.

We now describe how to extract the local magnetic field and crystal stress from the measured NV resonance frequencies. The NV ground-state Hamiltonian in the presence of stress and a static magnetic field is \cite{marcusStrainHam, galiSpinStrain, maletinskyStrainTerms}
\begin{equation}
\begin{split}
H_{\kappa} = & \left( D + M_{z,\kappa} \right) S_{z,\kappa}^2 + \gamma \vec{B} \cdot \vec{S}_{\kappa} \\
& + M_{x,\kappa} \left( S_{y,\kappa}^2 - S_{x,\kappa}^2 \right) \\
& + M_{y,\kappa} \left( S_{x,\kappa} S_{y,\kappa} + S_{y,\kappa} S_{x,\kappa} \right) \\
& + N_{x,\kappa} \left( S_{x,\kappa} S_{z,\kappa} + S_{z,\kappa} S_{x,\kappa} \right)  \\
& + N_{y,\kappa} \left( S_{y,\kappa} S_{z,\kappa} + S_{z,\kappa} S_{y,\kappa} \right).  \\
\end{split} 
\end{equation} 
Here, $D \approx 2870$ MHz is the zero-field splitting, $S_{i,\kappa}$ are the dimensionless spin-1 projection operators, $\gamma = 2.803$\e{4} MHz/T is the NV electronic gyromagnetic ratio, $\vec{B}$ is the magnetic field in the NV coordinate system, and $M_{i,\kappa}$ and $N_{i,\kappa}$ are terms related to the crystal stress and temperature. The indices $i = \{x,y,z\}$ represent the coordinate system for the particular NV orientation. We neglect the electric-field contributions to Eq.~1, as explained in the Supplementary Material \cite{suppl}. In addition, if $|\vec{B}| > 1$~mT, as is the case in this work, the contributions from the $\{ M_{x,\kappa}, M_{y,\kappa}, N_{x,\kappa}, N_{y,\kappa}\}$ terms are negligible, and Eq.~1 simplifies to
\begin{equation}
H_{\kappa} = (D + M_{z,\kappa}) S_{z,\kappa}^2 + \gamma \vec{B} \cdot \vec{S}_{\kappa}.
\label{eqHam}
\end{equation}
When $\vec{B}$ is aligned along the $z$-axis for one NV orientation, the Hamiltonian for the selected orientation reduces to
\begin{equation}
H_{\kappa} = (D + M_{z,\kappa}) S_{z,\kappa}^2 + \gamma B_{z} S_{z,\kappa},
\end{equation}
and the resonance frequencies are 
\begin{equation}
f_\pm = (D + M_{z,\kappa}) \pm \gamma B_{z}.
\end{equation}
Measuring $f_\pm$ yields the magnetic field projection $B_{z}$ and the $M_{z,\kappa}$ for that NV orientation. This measurement forms the basis of a sensing modality called Projection Magnetic Microscopy (PMM) \cite{QDM1ggg}, where we align the bias magnetic field along the $z$-axis of each NV orientation and record the associated resonance frequencies individually. An alternative sensing modality, called Vector Magnetic Microscopy (VMM) \cite{QDM1ggg}, allows us to determine $\vec{B}$ and all four $M_{z,\kappa}$ terms from a single measurement (Fig.~\ref{axesFig}b). In VMM, the selected bias magnetic field generates unique Zeeman splittings and non-overlapping ODMR spectra for each NV orientation. We extract the magnetic field components and  $M_{z,\kappa}$ values by fitting Eq.~2 for all four NV orientations. Both VMM and PMM yield the same $M_{z,\kappa}$ results; we detail advantages of each method in the Supplemental Material \cite{suppl}. We used VMM in this work to measure the four necessary $M_{z,\kappa}$ maps (which we refer to as ``NV $M_{z,\kappa}$ imaging") needed to reconstruct stress tensor elements for each pixel, as described below.

\begin{figure*}[ht]
\begin{center}
\begin{overpic}[width=0.98\textwidth]{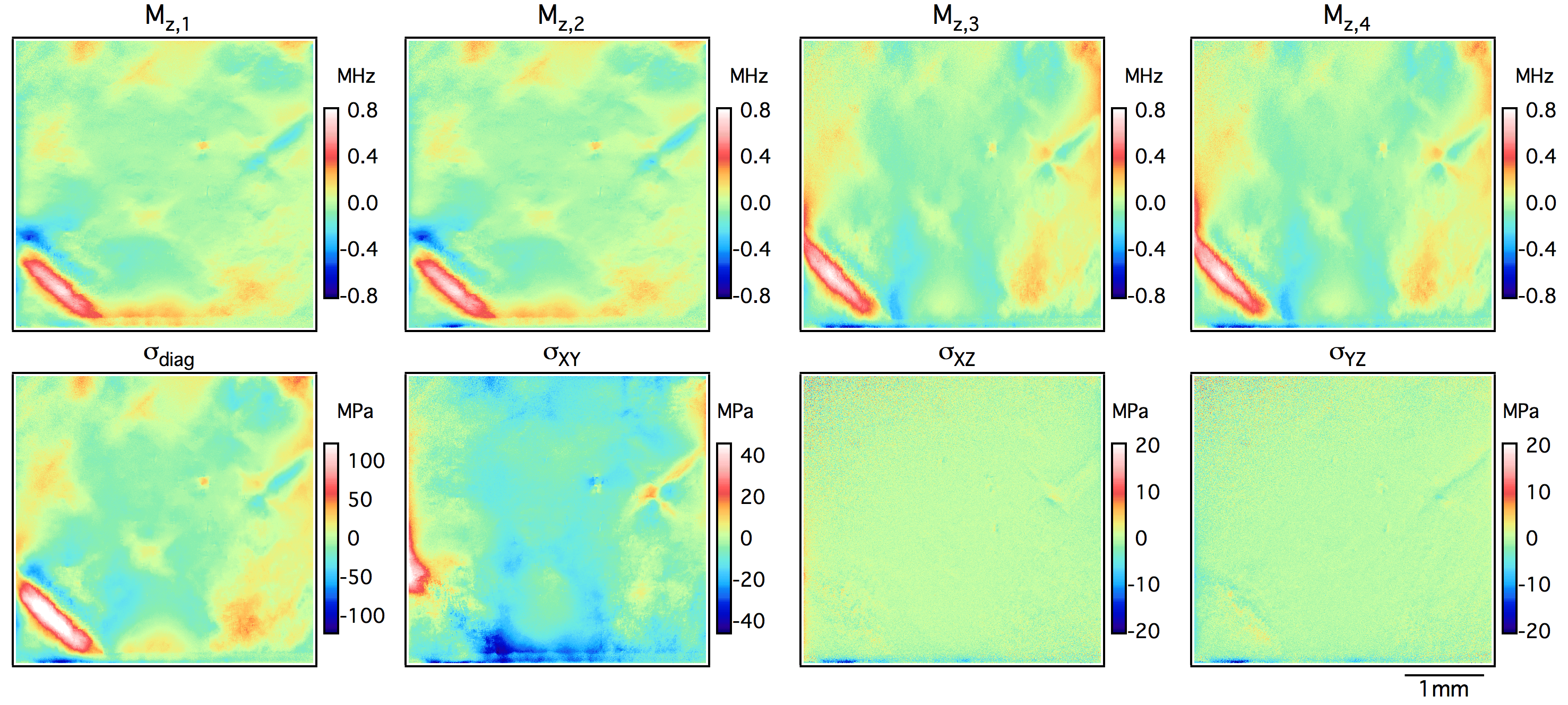}
\put(81,-4){\includegraphics[width=0.07\textwidth]{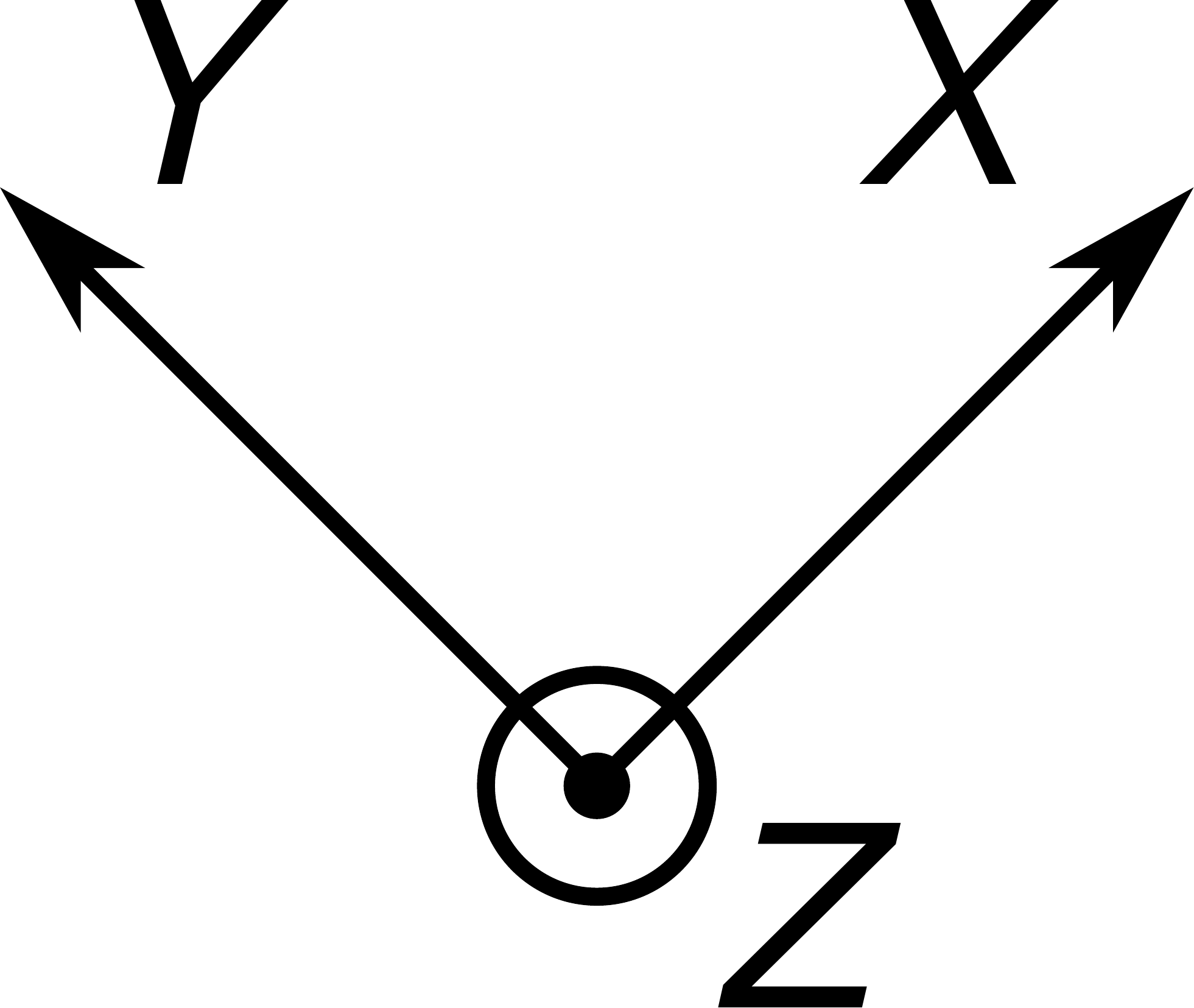}}
\end{overpic}
\end{center}
\caption{\label{MzToStressFig}
Example NV $M_{z,\kappa}$ and  $\{ \sigma_{\text{diag}}, \sigma_{XY}, \sigma_{XZ}, \sigma_{YZ} \}$  maps for Sample A. After measuring the $M_{z,\kappa}$ maps in the top row from the NV resonance frequencies, we calculate the stress tensor element maps in the bottom row using Eqs.~9-12. The diamond chip has high-stress and low-stress regions, and most of the $M_{z,\kappa}$ inhomogeneity comes from $\sigma_{\text{diag}}$ and $\sigma_{XY}$ stress terms. }
\end{figure*}

\begin{figure*}[ht]
\begin{center}
\begin{overpic}[width=0.98\textwidth]{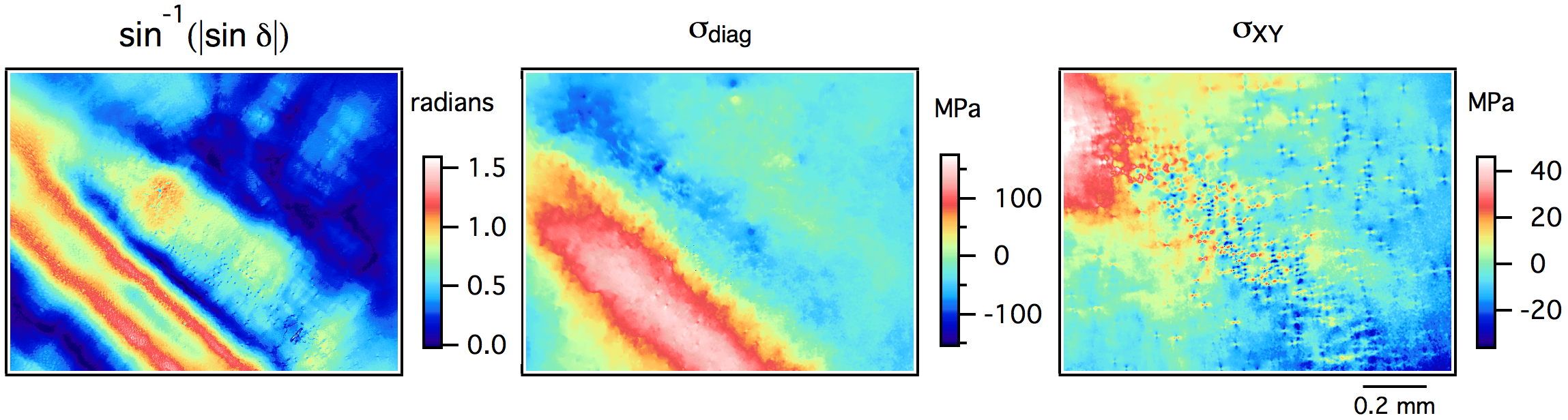}
\put(77,-4){\includegraphics[width=0.07\textwidth]{arrows.pdf}}
\end{overpic}
\end{center}
\caption{\label{birefringenceComparison}
Birefringence $\sin^{-1} |\sin{\delta}|$ and NV stress maps for the lower-left corner region of Sample A. Both techniques show similar phenomena, though the NV stress imaging maps are immune to the $\delta > \pi/2$ phase ambiguity, can resolve the petal-shaped defects localized in the NV layer, separate out strain phenomena into different stress tensor contributions, and predict how strain features affect the NV magnetic microscopy performance.}
\end{figure*}

\subsection{Stress tensor reconstruction}
In each pixel, the stress tensor components can be determined from the four $M_{z,\kappa}$ parameters, allowing us to generate a quantitative image of the local stress across the diamond. Following the derivations in Refs.~\cite{marcusStrainHam, galiSpinStrain, maletinskyStrainTerms}, we obtain 
\begin{align}
M_{z,1} & = a_1 \sigma_{\text{diag}} + 2 a_2 \left[ \sigma_{XY} + \sigma_{XZ} + \sigma_{YZ} \right],\\
M_{z,2} & = a_1 \sigma_{\text{diag}} + 2 a_2 \left[ \sigma_{XY} - \sigma_{XZ} - \sigma_{YZ} \right],\\
M_{z,3} & = a_1 \sigma_{\text{diag}} + 2 a_2 \left[ -\sigma_{XY} + \sigma_{XZ} - \sigma_{YZ} \right],\\
M_{z,4} & = a_1 \sigma_{\text{diag}} + 2 a_2 \left[ -\sigma_{XY} - \sigma_{XZ} + \sigma_{YZ} \right].
\end{align}
Here, $\{a_1, a_2\} = \{4.86, -3.7\}$ MHz/GPa are spin-stress coupling constants \cite{marcusStrainHam}, $\sigma_{ij}$ are elements of the 3$\times$3 stress tensor in GPa, and $\sigma_{\text{diag}} \equiv \sigma_{XX} + \sigma_{YY} + \sigma_{ZZ}$. The $\sigma_{ii}$ are normal stress terms, while $\sigma_{XY}$, $\sigma_{XZ}$, and $\sigma_{YZ}$ are shear stress terms. The $\sigma_{ij}$ are written in the diamond unit cell coordinate system $\{X,Y,Z\}$ (rather than the NV coordinate system $\{x,y,z\}$ for a given $\kappa$), and are felt by all four NV orientations. Each NV orientation exhibits the same $a_1 \sigma_{\text{diag}}$ contribution to $M_{z,\kappa}$. The $a_2$ contributions change as we transform the stress tensor for each of the four NV orientations.

Solving Eqs.~5-8 to extract $\sigma_{\text{diag}}$, $\sigma_{XY}$, $\sigma_{XZ}$, and $\sigma_{YZ}$ in each pixel yields
\begin{align}
\sigma_{\text{diag}} = & \frac{1}{4 a_1} \left[ M_{z,1} + M_{z,2} + M_{z,3} + M_{z,4} \right], \\
\sigma_{XY}          = & \frac{1}{8 a_2} \left[ M_{z,1} + M_{z,2} - M_{z,3} - M_{z,4} \right], \\
\sigma_{XZ}          = & \frac{1}{8 a_2} \left[ M_{z,1} - M_{z,2} + M_{z,3} - M_{z,4} \right], \\
\sigma_{YZ}          = & \frac{1}{8 a_2} \left[ M_{z,1} - M_{z,2} - M_{z,3} + M_{z,4} \right].
\end{align}
The measurements presented here are only sensitive to the total normal stress $\sigma_{\text{diag}}$ rather than the individual $\sigma_{ii}$ contributions \cite{marcusStrainHam}. A more sophisticated algorithm could use VMM spectra measured at several magnetic fields and keep all of the terms in Eq.~1 to obtain each $\sigma_{ii}$ separately. 

Since $M_{z,\kappa}$ and $\sigma_{\text{diag}}$ change with temperature as the diamond lattice constant changes, $M_{z,\kappa}$ and $\sigma_{\text{diag}}$ can only be evaluated up to an overall constant \cite{victor_Tdepend, marcus_temperature}. However, the shear stress terms should be unaffected by temperature changes, and thus shear stress images are absolute. For measurements acquired with 10 mK temperature stability, an NV $M_{z,\kappa}$ imager can determine $M_{z,\kappa}$ to about 1 kHz, or $\sim$0.1 MPa. As a further example, a 1 $\upmu \text{T}/\sqrt{\text{Hz}}$ magnetic sensitivity per pixel (28 kHz$/\sqrt{\text{Hz}}$ frequency sensitivity) corresponds to approximately 10 MPa$/\sqrt{\text{Hz}}$ stress sensitivity.

Figure \ref{MzToStressFig} shows the measured $M_{z,\kappa}$ maps and the resulting $\{ \sigma_{\text{diag}}, \sigma_{XY}, \sigma_{XZ}, \sigma_{YZ} \}$ maps for Sample A, illustrating a practical example of NV $M_{z,\kappa}$ and stress imaging. This diamond has a variety of strain features (their origins are described below), in addition to more homogeneous regions. For Sample A and most of the other diamond samples we investigated in this work, we found the shear stress inhomogeneity was greater in $\sigma_{XY}$ than in $\sigma_{XZ}$ or $\sigma_{YZ}$ \cite{suppl}. The $M_{z,\kappa}$ variations were usually due to $\sigma_{\text{diag}}$ and $\sigma_{XY}$ inhomogeneity in roughly equal amounts.

\section{Comparison with birefringence imaging}
Here we compare NV stress imaging to birefringence imaging, which is a prominent characterization tool in the diamond community \cite{FRIEL2009808, hoa2014}. In this work, both methods were implemented within the same optical microscope for a straightforward comparison (Fig.~\ref{setupFig}). Both the NV $M_{z,\kappa}$ terms and the diamond refractive index depend on crystal stress, but NV stress imaging more directly captures relevant information about stress inhomogeneity in the NV layer and its effects on NV sensing. This makes NV stress imaging the more appropriate tool for optimizing NV diamond samples for magnetic microscopy.

In a birefringent material, light with orthogonal polarizations transmitted through a sample of thickness $L$ accumulates a relative optical retardance phase  $\delta=\frac{2\pi}{\lambda}\Delta n L$, where $\lambda$ is the wavelength and $\Delta n$ is the difference in refractive indices for orthogonal polarizations. We used a rotating-linear-polarizer method, also known as Metripol, to extract $|\sin \delta|$ by probing the sample with  light of varying polarization angles \cite{glazer1996, kaminsky2007, suppl}. The measured transmission intensity $I_i$ for a given polarizer rotation angle $\alpha_i$ is
\begin{equation}
I_i=\frac{1}{2}I_0[1+\sin{2(\alpha_i-\phi)}\sin{\delta}].
\end{equation}
Here $I_0$ is the transmittance of a given pixel and $\phi$ is the retardance orientation angle. Sweeping $\alpha_i$ across $180\degree$ of polarization rotation allows us to determine $I_0$, $|\sin \delta|$, and $\phi$ \cite{suppl}.

Figure \ref{birefringenceComparison} shows a comparison between $\sin^{-1} |\sin{\delta}|$, $\sigma_{\text{diag}}$, and $\sigma_{XY}$ maps collected using birefringence and NV stress imaging with the same diamond field of view. Despite the general similarity in results between the two methods, there are some stark differences. The $\sigma_{XY}$ map shows petal-shaped strain features in the NV layer, whereas the birefringence map (which integrates phase retardance through the entire thickness) does not capture these fine details. Furthermore, the NV stress maps can distinguish that the diagonal stripe causing $M_{z,\kappa}$ inhomogeneity arises from $\sigma_{\text{diag}}$ stress, while the petal-shaped strain features are caused by $\sigma_{XY}$ stress. We can exploit such component-separated NV stress maps to investigate the sources and phenomenology of observed strain features.

Crystal strain and $\delta$ are linearly related through the diamond photo-elastic parameters \cite{howell2012Review, nye1957review, Ramachandran1947, hounsome2006}. However, this relationship typically assumes uniform stress over the optical path, meaning that the $\delta$ we measure is integrated over $L$ even though the strain may be localized to one layer. By comparison, the NV $M_{z,\kappa}$ technique provides  stress information localized to the NV layer, and converting from $M_{z,\kappa}$ to stress tensor elements is more straightforward.

Figure \ref{birefringenceComparison} illustrates an additional limitation for birefringence imaging. For high-strain regions, the integrated $\delta$ through the sample thickness may be greater than $\pi/2$, leading to ambiguity when calculating  stress from $|\sin \delta|$ since multiple $\delta$ values can yield the same $|\sin \delta|$. This occurs in the middle of the stripe feature in Fig.~\ref{birefringenceComparison}, where the reconstructed $\delta$ reaches its maximum value of $\pi/2$ before decreasing. NV stress imaging is not susceptible to this  ambiguity. The NV $\sigma_{\text{diag}}$ map instead shows that the stress amplitude increases to the middle of the stripe. Accounting for the extra $\sim\pi/4$ of phase accumulation in the birefringence map yields a maximum stress amplitude of $\sim$130 MPa, which is consistent with the 140 MPa maximum stress amplitude in the $\sigma_{\text{diag}}$ map \cite{suppl}. Despite the $|\sin \delta|$ ambiguity, the NV and birefringence methods yield consistent stress measurements.

\begin{figure}[ht]
\begin{center}
\begin{overpic}[width=0.9\columnwidth]{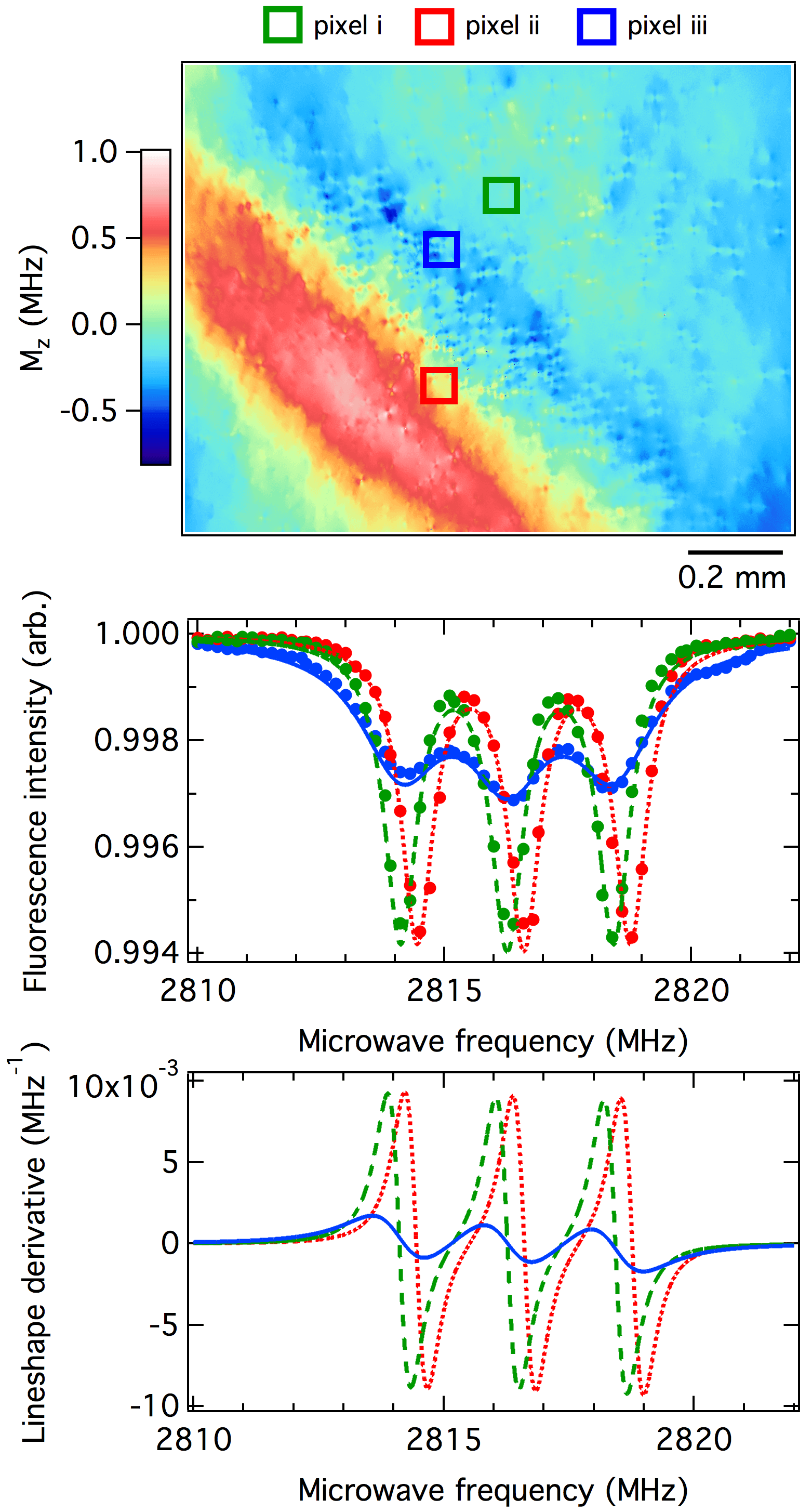}
\put(-2,98){\textsf{\Large a}}
\put(-2,62){\textsf{\Large b}}
\put(-2,31){\textsf{\Large c}}
\end{overpic}
\end{center}
\caption{\label{slopesExamplesFig}
(a) Zoomed-in $M_{z,\kappa}$ map (lower-left corner of Fig.~\ref{MzToStressFig}), showing the locations of the example pixels.
(b) Fitted ODMR spectra for example pixels (i), (ii), and (iii) (green, red, and blue, respectively). Each has varying $M_{z,\kappa}$ gradients and offsets.
(c) Derivatives $F'(f)$ for the ODMR lineshapes plotted in (b). Pixel (i) has the best $C/\Gamma$ slope and two-point responsivity, while pixel (ii) has poor $C/\Gamma$ slope and two-point responsivity due to the high $M_{z,\kappa}$ inhomogeneity in this pixel. Pixel (iii) has a good $C/\Gamma$ slope but a poor two-point responsivity, since the $M_{z,\kappa}$ offset means we probe this pixel at a suboptimal microwave frequencies compared to the others.}
\end{figure}

\section{Stress and NV magnetometry}
NV $M_{z,\kappa}$ inhomogeneity causes each NV in an ensemble to have different resonance frequencies, which reduces the magnetic sensitivity and degrades NV magnetometer performance \cite{P1DQ}. A useful NV-magnetometer figure of merit is the slope of the ODMR lineshape $|F'(f)|$, where $f$ is the probe-microwave frequency and $F'(f)$ is the derivative of the NV fluorescence intensity at frequency $f$ (Fig.~\ref{setupFig}b). The maximum $|F'(f)|$ slope is proportional to the quantity $ C / \Gamma$, where $C$ is the fluorescence contrast  and $\Gamma$ is the resonance linewidth \cite{kasperBookCh}.  $M_{z,\kappa}$ inhomogeneity reduces magnetic sensitivity by making the resonance lineshape broader, the contrast weaker, and thus the maximum slope $|F'(f)| \propto C/\Gamma$  smaller \cite{P1DQ}. 

For NV-diamond magnetometers that use fewer probe microwave frequencies for improved magnetic sensitivity, $M_{z,\kappa}$ inhomogeneity is even more detrimental. High-sensitivity magnetometers typically measure at two microwave frequencies (called the ``two-point method") instead of probing the full width of the ODMR lineshape (the ``full-sweep method") \cite{Glenn2018}. The two microwave frequencies are typically chosen to maximize the two-point responsivity (defined as the change in fluorescence per unit frequency shift of the NV resonance). If $M_{z,\kappa}$ varies substantially over the field of view, no pair of frequencies can be optimal for all NVs, resulting in decreased sensitivity for many pixels in the magnetic image. A larger variation in $M_{z,\kappa}$ across the ensemble also implies a narrower magnetic-field range before the NVs in some pixels fall out of resonance. As such, $M_{z,\kappa}$ inhomogeneity limits the field of view and dynamic range of high-sensitivity NV magnetic imagers. 

Figure \ref{slopesExamplesFig} shows a zoomed-in $M_{z,\kappa}$ map together with single-pixel ODMR spectra corresponding to regions of Sample A with different local strain properties. For example, one pixel shows a region with a low $M_{z,\kappa}$ gradient and offset from the mean (i); a second pixel shows a region with a low $M_{z,\kappa}$ gradient and a high $M_{z,\kappa}$ offset (ii); and a third pixel shows a region with a high $M_{z,\kappa}$ gradient and a low $M_{z,\kappa}$ offset (iii). These local strain conditions are caused by a $\sim$0.3 MHz $M_{z,\kappa}$ offset in the diagonal stripe and high $M_{z,\kappa}$ variation in the $\sim$30 $\upmu$m petal defects. Pixels (i) and (ii) have a comparable $C/\Gamma$ slope and therefore a comparable NV magnetic sensitivity when using the full-sweep method. However, when using the two-point method optimized for pixel (i), pixel (ii) will have a poor responsivity due to its large $M_{z,\kappa}$ offset. By comparison, pixel (iii) will exhibit poor performance with both methods. As these example pixels demonstrate, $M_{z,\kappa}$ inhomogeneity reduces the magnitude and uniformity of the magnetic sensitivity across an image.

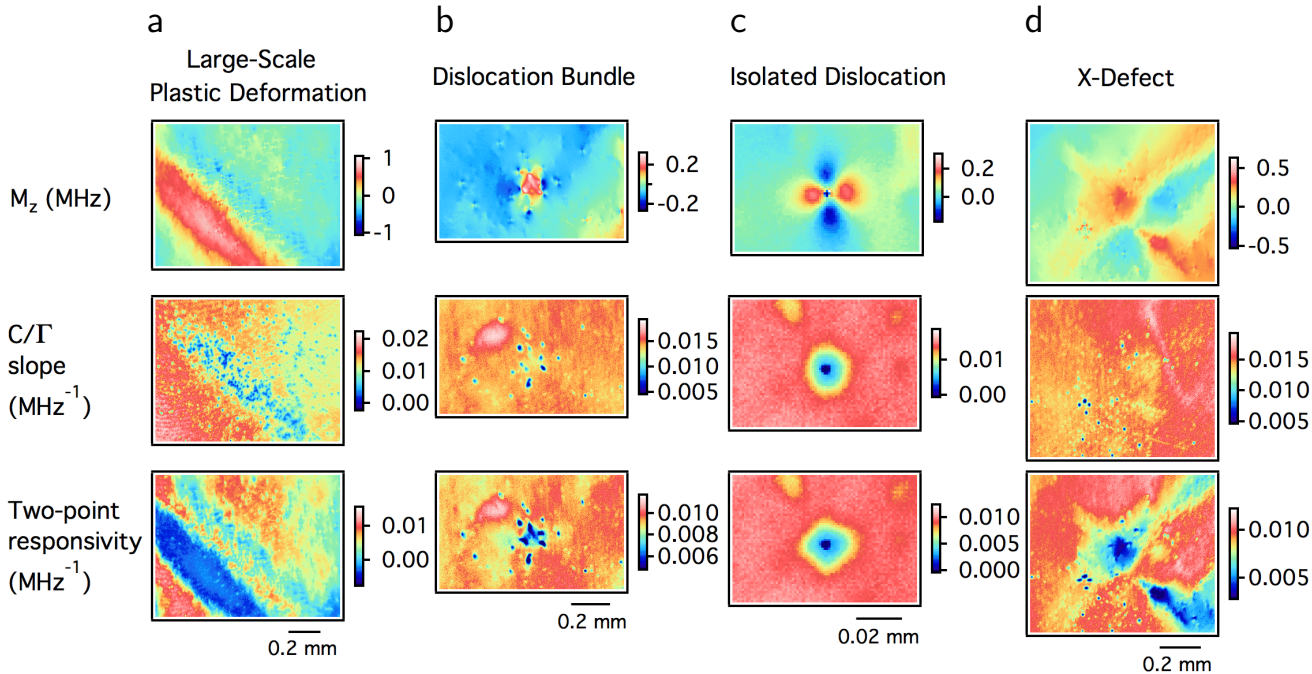
\begin{figure*}[ht]
\begin{center}
\begin{overpic}[width=0.98\textwidth]{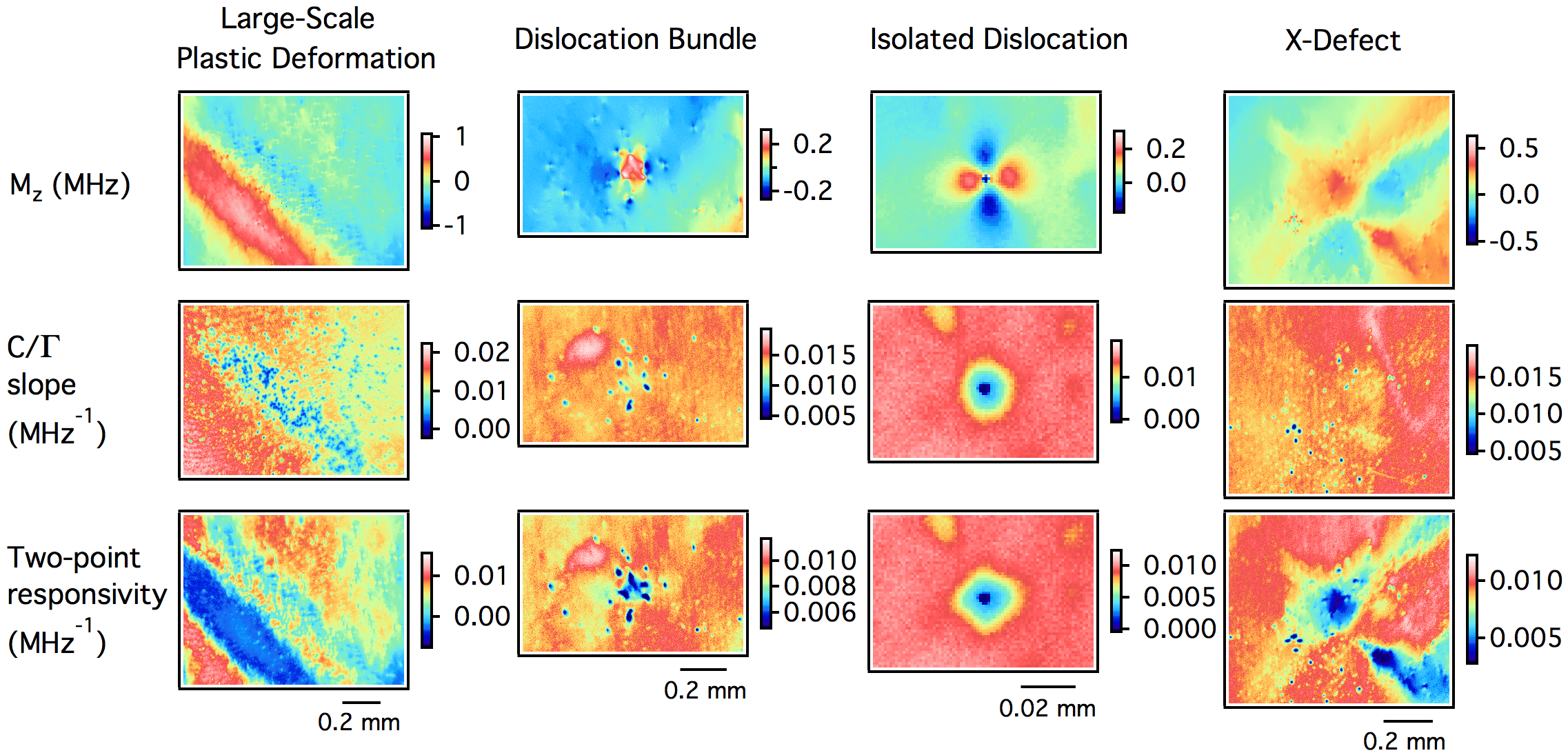}
\put(11,49){\textsf{\Large a}}
\put(33,49){\textsf{\Large b}}
\put(55.5,49){\textsf{\Large c}}
\put(78,49){\textsf{\Large d}}
\end{overpic}
\end{center}
\caption{\label{strainSurveyFig}
Comparisons between $M_{z,\kappa}$, $C/\Gamma$ slope, and two-point responsivity for common strain feature types found in Sample A. The $C/\Gamma$ and responsivity maps are related to the magnetic microscopy performance when using the full-sweep method and the two-point method, respectively. Note that the two-point responsivity is more susceptible to $M_{z,\kappa}$ inhomogeneity, while the full-sweep method can tolerate some range of $M_{z,\kappa}$ offsets. }
\end{figure*}

\section{Strain feature survey and effects on NV magnetometry}
We used NV $M_{z,\kappa}$ imaging to study and categorize different types of strain features in diamond samples. As shown in the regions highlighted in Fig.~\ref{strainSurveyFig}, different types of strain features have a variety of typical dimensions, $M_{z,\kappa}$ amplitudes and gradients, and stress tensor contributions. From our $M_{z,\kappa}$ maps, we categorized strain features into general types. We identified how each type impacts the $C/\Gamma$ slope and two-point responsivity. Here we concentrate on strain features observed in Sample A. Surveys of additional diamonds exhibiting similar phenomena are included in the Supplemental Material  \cite{suppl}.

Figure \ref{strainSurveyFig}a shows the same field of view as in Fig.~\ref{slopesExamplesFig}. The broad-scale plastic deformation in the diagonal stripe is perhaps associated with the lower-left corner of the diamond sample, as high stress is common at  sharp corners, edges, and fractures. The stress from the diagonal stripe is largely $\sigma_{\text{diag}}$ stress, causing millimeter-scale $M_{z,\kappa}$ gradients, resulting in a wide span in NV resonance frequencies ($\sim$1 MHz). As anticipated, the diagonal stripe spoils the two-point responsivity while the full-sweep $C/\Gamma$ slope is largely unaffected. In this example the $M_{z,\kappa}$ span is large enough to cause a negative responsivity in the diagonal stripe, as the resonance frequency is offset far enough that one of the probe frequencies is on the opposite side of its resonance peak.

The 20-30 $\upmu$m petal-shaped strain defects in Fig.~\ref{strainSurveyFig}a and Fig.~\ref{birefringenceComparison} are caused by lattice dislocations that can form on top of the seed crystal during homoepitaxial growth, as studied in previous work \cite{twitchenXray, shikataXray, martineauXray, FRIEL2009808, hoa2014}. The three types of lattice dislocations (edge, screw, and mixed dislocations) contribute to different crystal stress terms \cite{hullBacon}. The petal features appear most strongly in the $\sigma_{\text{diag}}$ and $\sigma_{XY}$ maps (and to a lesser degree in the $\sigma_{XZ}$ and $\sigma_{YZ}$ maps), which suggests that the petal-shaped strain features we observed are predominantly caused by edge and mixed dislocations. 

Figure \ref{strainSurveyFig}b shows a $\sim$200 $\upmu$m strain feature (likely caused by a dislocation bundle), surrounded by smaller petal-shaped defects. From birefringence imaging, we know that such strain features are typically edge dislocations (with $\sigma_{\text{diag}}$  and $\sigma_{XY}$ stress). They often have four quadrants with lines emanating from the center along the [001] and [010] directions, and are a few hundred micrometers across \cite{Pinto2009}. As shown in the Supplemental Material \cite{suppl}, the birefringence map displays  lobes associated with the strain feature in Fig.~\ref{strainSurveyFig}b, with the expected orientation. The lobes appearing in the $\sigma_{\text{diag}}$  and $\sigma_{XY}$ NV stress maps are rotated by 45$\degree$. These characteristics lead us to conclude that this strain feature is a dislocation bundle. For this particular strain feature, the range of $M_{z,\kappa}$ values is narrow enough that it has only a minor effect on NV magnetometry performance for both the full-sweep and two-point methods.

Figure \ref{strainSurveyFig}c shows a prominent $\sim$30 $\upmu$m dislocation strain feature. Here the single-pixel $M_{z,\kappa}$ gradients are substantial enough to spoil the NV magnetic sensitivity of both methods. Severe $M_{z,\kappa}$ gradients also interfere when fitting the ODMR spectra to a Lorentzian lineshape model, introducing  systematic errors in the extracted resonance frequencies. Such errors can produce false features in NV magnetic images \cite{suppl}.

Figure \ref{strainSurveyFig}d shows a $\sim$0.8 mm X-shaped strain feature. Though visually most similar to the petal-shaped strain features discussed above, X-shaped strain features are larger, display sharp edges pointing along the diamond [100] and [010] directions, and have no lobe structures. The X-shaped strain features also exhibit mainly $\sigma_{\text{diag}}$  and $\sigma_{XY}$ stress (like an edge dislocation), whereas the $\sigma_{XZ}$ and $\sigma_{YZ}$ values are nearly zero. Despite the similarities to the previously-discussed strain features, the origin of the X-shaped strain features remains under investigation. They mainly affect the two-point responsivity for NV magnetometry, whereas the full-sweep $C/\Gamma$ slope is mostly immune.

\section{Summary and outlook}
We presented a method for quantitative stress imaging in diamond with micrometer spatial resolution and millimeter field of view using a layer of NV centers. We compared NV stress imaging to the more traditional birefringence imaging method, implemented in the same experimental setup, and found quantitative and qualitative consistency. NV $M_{z,\kappa}$ imaging offers a straightforward way to reconstruct stress tensor elements within a diamond sample and provides a more direct measure of how the strain features affect NV magnetic imaging. NV $M_{z,\kappa}$ imaging is therefore a useful tool to support NV magnetic microscopy and other diamond applications that rely on crystal homogeneity for optimal performance. 

To further improve the NV $M_{z,\kappa}$ imaging method, one can implement NV sensitivity and resolution enhancements. For example, one can boost the sensitivity by implementing a double-quantum Ramsey spectroscopy protocol, creating a superposition of the NV $m_s = \pm1$ magnetic sublevels. This doubles the $M_{z,\kappa}$ part of Eq.~\ref{eqHam} and cancels the magnetic contribution \cite{P1DQ, ToyliTemperature}. NV $M_{z,\kappa}$ imaging with double-quantum Ramsey spectroscopy should be beneficial for NV layers where the magnetic field inhomogeneity dominates the $M_{z,\kappa}$ inhomogeneity. Furthermore, for specific applications, one can perform additional measurements to disentangle the $\sigma_{ii}$ normal stress terms. Finally, one can employ NV super-resolution techniques to map the stress tensor components with a resolution beyond the optical diffraction limit \cite{STED1, JCsuperres}.

Looking to future diamond applications for particle physics, diamond stress characterization is important for the recently-proposed diamond directional weakly-interacting massive particle (WIMP) detector \cite{nvWIMPdetector}. This approach aims to use NV centers to image the stress created by $\sim$100 nm tracks from recoiling carbon nuclei. Mapping the intrinsic $M_{z,\kappa}$ and stress inhomogeneity is a first step to exploring the feasibility of directional WIMP detection with NVs. In particular, since $\sigma_{XZ}$ and $\sigma_{YZ}$ stress are typically smaller than $\sigma_{\text{diag}}$ and $\sigma_{XY}$ stress, detecting deviations in $\sigma_{XZ}$ or $\sigma_{YZ}$ may exhibit a larger signal-to-background ratio. Anticipated next steps include NV $M_{z,\kappa}$ imaging with higher spatial resolution ($<$1 $\upmu$m) and variable depth; cataloging the $M_{z,\kappa}$ distribution from many individual NV centers in a low-density bulk sample (ppb NV density); investigating hybrid-sensor schemes (such as a combined cathodoluminescence/$M_{z,\kappa}$ method) to rapidly survey diamond chips for damaged voxels; and imaging the recoil tracks from implanted $^{12}$C nuclei.

\section{Acknowledgements}
We thank Marcus Doherty and Adam Gali for help with clarifying the crystal stress contributions to the NV Hamiltonian and Michel Mermoux for insights on stress tensor extraction methods. We thank Patrick Scheidegger for assistance adopting GPU-accelerated data analysis and Connor Finnerty for assistance in automating the birefringence imaging. We also thank Abdelghani Laraoui, Andrew Mounce, and David Phillips for providing feedback on the manuscript. While preparing this paper we became aware of a preprint \cite{hollenbergStrain} which presents a similar NV stress mapping scheme, though our work studies naturally-formed defects to optimize NV magnetic imaging. This work was supported by DOE award DE‐SC0019396; DARPA DRINQS award D18AC00033; Air Force Office of Scientific Research award FA9550-17-1-0371; and NSF awards PHY-1504610 and EAR 1647504. This work was performed in part at the Harvard Center for Nanoscale Systems (CNS), a member of the National Nanotechnology Coordinated Infrastructure Network (NNCI), which is supported by the National Science Foundation under NSF award no.~1541959. We thank Edward Soucy, Brett Graham, and the Harvard Center for Brain Science for technical support and fabrication assistance.


\begin{thebibliography}{37}%
\makeatletter
\providecommand \@ifxundefined [1]{%
 \@ifx{#1\undefined}
}%
\providecommand \@ifnum [1]{%
 \ifnum #1\expandafter \@firstoftwo
 \else \expandafter \@secondoftwo
 \fi
}%
\providecommand \@ifx [1]{%
 \ifx #1\expandafter \@firstoftwo
 \else \expandafter \@secondoftwo
 \fi
}%
\providecommand \natexlab [1]{#1}%
\providecommand \enquote  [1]{``#1''}%
\providecommand \bibnamefont  [1]{#1}%
\providecommand \bibfnamefont [1]{#1}%
\providecommand \citenamefont [1]{#1}%
\providecommand \href@noop [0]{\@secondoftwo}%
\providecommand \href [0]{\begingroup \@sanitize@url \@href}%
\providecommand \@href[1]{\@@startlink{#1}\@@href}%
\providecommand \@@href[1]{\endgroup#1\@@endlink}%
\providecommand \@sanitize@url [0]{\catcode `\\12\catcode `\$12\catcode
  `\&12\catcode `\#12\catcode `\^12\catcode `\_12\catcode `\%12\relax}%
\providecommand \@@startlink[1]{}%
\providecommand \@@endlink[0]{}%
\providecommand \url  [0]{\begingroup\@sanitize@url \@url }%
\providecommand \@url [1]{\endgroup\@href {#1}{\urlprefix }}%
\providecommand \urlprefix  [0]{URL }%
\providecommand \Eprint [0]{\href }%
\providecommand \doibase [0]{http://dx.doi.org/}%
\providecommand \selectlanguage [0]{\@gobble}%
\providecommand \bibinfo  [0]{\@secondoftwo}%
\providecommand \bibfield  [0]{\@secondoftwo}%
\providecommand \translation [1]{[#1]}%
\providecommand \BibitemOpen [0]{}%
\providecommand \bibitemStop [0]{}%
\providecommand \bibitemNoStop [0]{.\EOS\space}%
\providecommand \EOS [0]{\spacefactor3000\relax}%
\providecommand \BibitemShut  [1]{\csname bibitem#1\endcsname}%
\let\auto@bib@innerbib\@empty
\bibitem [{\citenamefont {Schlussel}\ \emph {et~al.}(2018)\citenamefont
  {Schlussel}, \citenamefont {Lenz}, \citenamefont {Rohner}, \citenamefont
  {Bar-Haim}, \citenamefont {Bougas}, \citenamefont {Groswasser}, \citenamefont
  {Kieschnick}, \citenamefont {Rozenberg}, \citenamefont {Thiel}, \citenamefont
  {Waxman}, \citenamefont {Meijer}, \citenamefont {Maletinsky}, \citenamefont
  {Budker},\ and\ \citenamefont {Folman}}]{heziVortices}%
  \BibitemOpen
  \bibfield  {author} {\bibinfo {author} {\bibfnamefont {Y.}~\bibnamefont
  {Schlussel}}, \bibinfo {author} {\bibfnamefont {T.}~\bibnamefont {Lenz}},
  \bibinfo {author} {\bibfnamefont {D.}~\bibnamefont {Rohner}}, \bibinfo
  {author} {\bibfnamefont {Y.}~\bibnamefont {Bar-Haim}}, \bibinfo {author}
  {\bibfnamefont {L.}~\bibnamefont {Bougas}}, \bibinfo {author} {\bibfnamefont
  {D.}~\bibnamefont {Groswasser}}, \bibinfo {author} {\bibfnamefont
  {M.}~\bibnamefont {Kieschnick}}, \bibinfo {author} {\bibfnamefont
  {E.}~\bibnamefont {Rozenberg}}, \bibinfo {author} {\bibfnamefont
  {L.}~\bibnamefont {Thiel}}, \bibinfo {author} {\bibfnamefont
  {A.}~\bibnamefont {Waxman}}, \bibinfo {author} {\bibfnamefont
  {J.}~\bibnamefont {Meijer}}, \bibinfo {author} {\bibfnamefont
  {P.}~\bibnamefont {Maletinsky}}, \bibinfo {author} {\bibfnamefont
  {D.}~\bibnamefont {Budker}}, \ and\ \bibinfo {author} {\bibfnamefont
  {R.}~\bibnamefont {Folman}},\ }\href {\doibase
  10.1103/PhysRevApplied.10.034032} {\bibfield  {journal} {\bibinfo  {journal}
  {Phys. Rev. Applied}\ }\textbf {\bibinfo {volume} {10}},\ \bibinfo {pages}
  {034032} (\bibinfo {year} {2018})}\BibitemShut {NoStop}%
\bibitem [{\citenamefont {Tetienne}\ \emph {et~al.}(2017)\citenamefont
  {Tetienne}, \citenamefont {Dontschuk}, \citenamefont {Broadway},
  \citenamefont {Stacey}, \citenamefont {Simpson},\ and\ \citenamefont
  {Hollenberg}}]{tetienneGraphene}%
  \BibitemOpen
  \bibfield  {author} {\bibinfo {author} {\bibfnamefont {J.-P.}\ \bibnamefont
  {Tetienne}}, \bibinfo {author} {\bibfnamefont {N.}~\bibnamefont {Dontschuk}},
  \bibinfo {author} {\bibfnamefont {D.~A.}\ \bibnamefont {Broadway}}, \bibinfo
  {author} {\bibfnamefont {A.}~\bibnamefont {Stacey}}, \bibinfo {author}
  {\bibfnamefont {D.~A.}\ \bibnamefont {Simpson}}, \ and\ \bibinfo {author}
  {\bibfnamefont {L.~C.~L.}\ \bibnamefont {Hollenberg}},\ }\href
  {http://advances.sciencemag.org/content/3/4/e1602429} {\bibfield  {journal}
  {\bibinfo  {journal} {Science Advances}\ }\textbf {\bibinfo {volume} {3}}
  (\bibinfo {year} {2017})}\BibitemShut {NoStop}%
\bibitem [{\citenamefont {Glenn}\ \emph {et~al.}(2017)\citenamefont {Glenn},
  \citenamefont {Fu}, \citenamefont {Kehayias}, \citenamefont {Le~Sage},
  \citenamefont {Lima}, \citenamefont {Weiss},\ and\ \citenamefont
  {Walsworth}}]{QDM1ggg}%
  \BibitemOpen
  \bibfield  {author} {\bibinfo {author} {\bibfnamefont {D.~R.}\ \bibnamefont
  {Glenn}}, \bibinfo {author} {\bibfnamefont {R.~R.}\ \bibnamefont {Fu}},
  \bibinfo {author} {\bibfnamefont {P.}~\bibnamefont {Kehayias}}, \bibinfo
  {author} {\bibfnamefont {D.}~\bibnamefont {Le~Sage}}, \bibinfo {author}
  {\bibfnamefont {E.~A.}\ \bibnamefont {Lima}}, \bibinfo {author}
  {\bibfnamefont {B.~P.}\ \bibnamefont {Weiss}}, \ and\ \bibinfo {author}
  {\bibfnamefont {R.~L.}\ \bibnamefont {Walsworth}},\ }\href {\doibase
  10.1002/2017GC006946} {\bibfield  {journal} {\bibinfo  {journal}
  {Geochemistry, Geophysics, Geosystems}\ }\textbf {\bibinfo {volume} {18}},\
  \bibinfo {pages} {3254} (\bibinfo {year} {2017})}\BibitemShut {NoStop}%
\bibitem [{\citenamefont {Le~Sage}\ \emph {et~al.}(2013)\citenamefont
  {Le~Sage}, \citenamefont {Arai}, \citenamefont {Glenn}, \citenamefont
  {DeVience}, \citenamefont {Pham}, \citenamefont {Rahn-Lee}, \citenamefont
  {Lukin}, \citenamefont {Yacoby}, \citenamefont {Komeili},\ and\ \citenamefont
  {Walsworth}}]{nv_bacteria}%
  \BibitemOpen
  \bibfield  {author} {\bibinfo {author} {\bibfnamefont {D.}~\bibnamefont
  {Le~Sage}}, \bibinfo {author} {\bibfnamefont {K.}~\bibnamefont {Arai}},
  \bibinfo {author} {\bibfnamefont {D.~R.}\ \bibnamefont {Glenn}}, \bibinfo
  {author} {\bibfnamefont {S.~J.}\ \bibnamefont {DeVience}}, \bibinfo {author}
  {\bibfnamefont {L.~M.}\ \bibnamefont {Pham}}, \bibinfo {author}
  {\bibfnamefont {L.}~\bibnamefont {Rahn-Lee}}, \bibinfo {author}
  {\bibfnamefont {M.~D.}\ \bibnamefont {Lukin}}, \bibinfo {author}
  {\bibfnamefont {A.}~\bibnamefont {Yacoby}}, \bibinfo {author} {\bibfnamefont
  {A.}~\bibnamefont {Komeili}}, \ and\ \bibinfo {author} {\bibfnamefont
  {R.~L.}\ \bibnamefont {Walsworth}},\ }\href
  {http://dx.doi.org/10.1038/nature12072} {\bibfield  {journal} {\bibinfo
  {journal} {Nature}\ }\textbf {\bibinfo {volume} {496}},\ \bibinfo {pages}
  {486} (\bibinfo {year} {2013})}\BibitemShut {NoStop}%
\bibitem [{\citenamefont {Fescenko}\ \emph {et~al.}(2019)\citenamefont
  {Fescenko}, \citenamefont {Laraoui}, \citenamefont {Smits}, \citenamefont
  {Mosavian}, \citenamefont {Kehayias}, \citenamefont {Seto}, \citenamefont
  {Bougas}, \citenamefont {Jarmola},\ and\ \citenamefont {Acosta}}]{hemozoin}%
  \BibitemOpen
  \bibfield  {author} {\bibinfo {author} {\bibfnamefont {I.}~\bibnamefont
  {Fescenko}}, \bibinfo {author} {\bibfnamefont {A.}~\bibnamefont {Laraoui}},
  \bibinfo {author} {\bibfnamefont {J.}~\bibnamefont {Smits}}, \bibinfo
  {author} {\bibfnamefont {N.}~\bibnamefont {Mosavian}}, \bibinfo {author}
  {\bibfnamefont {P.}~\bibnamefont {Kehayias}}, \bibinfo {author}
  {\bibfnamefont {J.}~\bibnamefont {Seto}}, \bibinfo {author} {\bibfnamefont
  {L.}~\bibnamefont {Bougas}}, \bibinfo {author} {\bibfnamefont
  {A.}~\bibnamefont {Jarmola}}, \ and\ \bibinfo {author} {\bibfnamefont
  {V.~M.}\ \bibnamefont {Acosta}},\ }\href {\doibase
  10.1103/PhysRevApplied.11.034029} {\bibfield  {journal} {\bibinfo  {journal}
  {Phys. Rev. Applied}\ }\textbf {\bibinfo {volume} {11}},\ \bibinfo {pages}
  {034029} (\bibinfo {year} {2019})}\BibitemShut {NoStop}%
\bibitem [{\citenamefont {Bauch}\ \emph {et~al.}(2018)\citenamefont {Bauch},
  \citenamefont {Hart}, \citenamefont {Schloss}, \citenamefont {Turner},
  \citenamefont {Barry}, \citenamefont {Kehayias}, \citenamefont {Singh},\ and\
  \citenamefont {Walsworth}}]{P1DQ}%
  \BibitemOpen
  \bibfield  {author} {\bibinfo {author} {\bibfnamefont {E.}~\bibnamefont
  {Bauch}}, \bibinfo {author} {\bibfnamefont {C.~A.}\ \bibnamefont {Hart}},
  \bibinfo {author} {\bibfnamefont {J.~M.}\ \bibnamefont {Schloss}}, \bibinfo
  {author} {\bibfnamefont {M.~J.}\ \bibnamefont {Turner}}, \bibinfo {author}
  {\bibfnamefont {J.~F.}\ \bibnamefont {Barry}}, \bibinfo {author}
  {\bibfnamefont {P.}~\bibnamefont {Kehayias}}, \bibinfo {author}
  {\bibfnamefont {S.}~\bibnamefont {Singh}}, \ and\ \bibinfo {author}
  {\bibfnamefont {R.~L.}\ \bibnamefont {Walsworth}},\ }\href {\doibase
  10.1103/PhysRevX.8.031025} {\bibfield  {journal} {\bibinfo  {journal} {Phys.
  Rev. X}\ }\textbf {\bibinfo {volume} {8}},\ \bibinfo {pages} {031025}
  (\bibinfo {year} {2018})}\BibitemShut {NoStop}%
\bibitem [{\citenamefont {Barry}\ \emph {et~al.}(2016)\citenamefont {Barry},
  \citenamefont {Turner}, \citenamefont {Schloss}, \citenamefont {Glenn},
  \citenamefont {Song}, \citenamefont {Lukin}, \citenamefont {Park},\ and\
  \citenamefont {Walsworth}}]{barryNeurons}%
  \BibitemOpen
  \bibfield  {author} {\bibinfo {author} {\bibfnamefont {J.~F.}\ \bibnamefont
  {Barry}}, \bibinfo {author} {\bibfnamefont {M.~J.}\ \bibnamefont {Turner}},
  \bibinfo {author} {\bibfnamefont {J.~M.}\ \bibnamefont {Schloss}}, \bibinfo
  {author} {\bibfnamefont {D.~R.}\ \bibnamefont {Glenn}}, \bibinfo {author}
  {\bibfnamefont {Y.}~\bibnamefont {Song}}, \bibinfo {author} {\bibfnamefont
  {M.~D.}\ \bibnamefont {Lukin}}, \bibinfo {author} {\bibfnamefont
  {H.}~\bibnamefont {Park}}, \ and\ \bibinfo {author} {\bibfnamefont {R.~L.}\
  \bibnamefont {Walsworth}},\ }\href@noop {} {\bibfield  {journal} {\bibinfo
  {journal} {Proceedings of the National Academy of Sciences}\ } (\bibinfo
  {year} {2016})}\BibitemShut {NoStop}%
\bibitem [{\citenamefont {Jensen}\ \emph {et~al.}(2014)\citenamefont {Jensen},
  \citenamefont {Leefer}, \citenamefont {Jarmola}, \citenamefont {Dumeige},
  \citenamefont {Acosta}, \citenamefont {Kehayias}, \citenamefont {Patton},\
  and\ \citenamefont {Budker}}]{kasper_IR}%
  \BibitemOpen
  \bibfield  {author} {\bibinfo {author} {\bibfnamefont {K.}~\bibnamefont
  {Jensen}}, \bibinfo {author} {\bibfnamefont {N.}~\bibnamefont {Leefer}},
  \bibinfo {author} {\bibfnamefont {A.}~\bibnamefont {Jarmola}}, \bibinfo
  {author} {\bibfnamefont {Y.}~\bibnamefont {Dumeige}}, \bibinfo {author}
  {\bibfnamefont {V.~M.}\ \bibnamefont {Acosta}}, \bibinfo {author}
  {\bibfnamefont {P.}~\bibnamefont {Kehayias}}, \bibinfo {author}
  {\bibfnamefont {B.}~\bibnamefont {Patton}}, \ and\ \bibinfo {author}
  {\bibfnamefont {D.}~\bibnamefont {Budker}},\ }\href {\doibase
  10.1103/PhysRevLett.112.160802} {\bibfield  {journal} {\bibinfo  {journal}
  {Phys. Rev. Lett.}\ }\textbf {\bibinfo {volume} {112}},\ \bibinfo {pages}
  {160802} (\bibinfo {year} {2014})}\BibitemShut {NoStop}%
\bibitem [{\citenamefont {Spear}\ and\ \citenamefont
  {Dismukes}(1994)}]{synthDiamondBook}%
  \BibitemOpen
  \bibfield  {author} {\bibinfo {author} {\bibfnamefont {K.~E.}\ \bibnamefont
  {Spear}}\ and\ \bibinfo {author} {\bibfnamefont {J.~P.}\ \bibnamefont
  {Dismukes}},\ }\href@noop {} {\emph {\bibinfo {title} {Synthetic Diamond:
  Emerging CVD Science and Technology}}}\ (\bibinfo  {publisher} {Wiley},\
  \bibinfo {year} {1994})\BibitemShut {NoStop}%
\bibitem [{sup()}]{suppl}%
  \BibitemOpen
  \href@noop {} {}\bibinfo {note} {Additional details are included in the
  supplemental material.}\BibitemShut {Stop}%
\bibitem [{\citenamefont {Friel}\ \emph {et~al.}(2009)\citenamefont {Friel},
  \citenamefont {Clewes}, \citenamefont {Dhillon}, \citenamefont {Perkins},
  \citenamefont {Twitchen},\ and\ \citenamefont {Scarsbrook}}]{FRIEL2009808}%
  \BibitemOpen
  \bibfield  {author} {\bibinfo {author} {\bibfnamefont {I.}~\bibnamefont
  {Friel}}, \bibinfo {author} {\bibfnamefont {S.}~\bibnamefont {Clewes}},
  \bibinfo {author} {\bibfnamefont {H.}~\bibnamefont {Dhillon}}, \bibinfo
  {author} {\bibfnamefont {N.}~\bibnamefont {Perkins}}, \bibinfo {author}
  {\bibfnamefont {D.}~\bibnamefont {Twitchen}}, \ and\ \bibinfo {author}
  {\bibfnamefont {G.}~\bibnamefont {Scarsbrook}},\ }\href {\doibase
  https://doi.org/10.1016/j.diamond.2009.01.013} {\bibfield  {journal}
  {\bibinfo  {journal} {Diamond and Related Materials}\ }\textbf {\bibinfo
  {volume} {18}},\ \bibinfo {pages} {808 } (\bibinfo {year}
  {2009})}\BibitemShut {NoStop}%
\bibitem [{\citenamefont {Crisci}\ \emph {et~al.}(2011)\citenamefont {Crisci},
  \citenamefont {Baillet}, \citenamefont {Mermoux}, \citenamefont {Bogdan},
  \citenamefont {Nesl\'adek},\ and\ \citenamefont {Haenen}}]{crisci2011}%
  \BibitemOpen
  \bibfield  {author} {\bibinfo {author} {\bibfnamefont {A.}~\bibnamefont
  {Crisci}}, \bibinfo {author} {\bibfnamefont {F.}~\bibnamefont {Baillet}},
  \bibinfo {author} {\bibfnamefont {M.}~\bibnamefont {Mermoux}}, \bibinfo
  {author} {\bibfnamefont {G.}~\bibnamefont {Bogdan}}, \bibinfo {author}
  {\bibfnamefont {M.}~\bibnamefont {Nesl\'adek}}, \ and\ \bibinfo {author}
  {\bibfnamefont {K.}~\bibnamefont {Haenen}},\ }\href {\doibase
  10.1002/pssa.201100039} {\bibfield  {journal} {\bibinfo  {journal} {physica
  status solidi (a)}\ }\textbf {\bibinfo {volume} {208}},\ \bibinfo {pages}
  {2038} (\bibinfo {year} {2011})}\BibitemShut {NoStop}%
\bibitem [{\citenamefont {Hanley}\ \emph {et~al.}(1977)\citenamefont {Hanley},
  \citenamefont {Kiflawi},\ and\ \citenamefont {Lang}}]{hanleyCath}%
  \BibitemOpen
  \bibfield  {author} {\bibinfo {author} {\bibfnamefont {P.~L.}\ \bibnamefont
  {Hanley}}, \bibinfo {author} {\bibfnamefont {I.}~\bibnamefont {Kiflawi}}, \
  and\ \bibinfo {author} {\bibfnamefont {A.~R.}\ \bibnamefont {Lang}},\
  }\href@noop {} {\bibfield  {journal} {\bibinfo  {journal} {Proceedings of the
  Royal Society of London. Series A, Mathematical and Physical Sciences}\
  }\textbf {\bibinfo {volume} {284}},\ \bibinfo {pages} {329} (\bibinfo {year}
  {1977})}\BibitemShut {NoStop}%
\bibitem [{\citenamefont {Broadway}\ \emph {et~al.}(2018)\citenamefont
  {Broadway}, \citenamefont {Johnson}, \citenamefont {Barson}, \citenamefont
  {Lillie}, \citenamefont {Dontschuk}, \citenamefont {McCloskey}, \citenamefont
  {Tsai}, \citenamefont {Teraji}, \citenamefont {Simpson}, \citenamefont
  {Stacey}, \citenamefont {McCallum}, \citenamefont {Bradby}, \citenamefont
  {Doherty}, \citenamefont {Hollenberg},\ and\ \citenamefont
  {Tetienne}}]{hollenbergStrain}%
  \BibitemOpen
  \bibfield  {author} {\bibinfo {author} {\bibfnamefont {D.~A.}\ \bibnamefont
  {Broadway}}, \bibinfo {author} {\bibfnamefont {B.~C.}\ \bibnamefont
  {Johnson}}, \bibinfo {author} {\bibfnamefont {M.~S.~J.}\ \bibnamefont
  {Barson}}, \bibinfo {author} {\bibfnamefont {S.~E.}\ \bibnamefont {Lillie}},
  \bibinfo {author} {\bibfnamefont {N.}~\bibnamefont {Dontschuk}}, \bibinfo
  {author} {\bibfnamefont {D.~J.}\ \bibnamefont {McCloskey}}, \bibinfo {author}
  {\bibfnamefont {A.}~\bibnamefont {Tsai}}, \bibinfo {author} {\bibfnamefont
  {T.}~\bibnamefont {Teraji}}, \bibinfo {author} {\bibfnamefont {D.~A.}\
  \bibnamefont {Simpson}}, \bibinfo {author} {\bibfnamefont {A.}~\bibnamefont
  {Stacey}}, \bibinfo {author} {\bibfnamefont {J.~C.}\ \bibnamefont
  {McCallum}}, \bibinfo {author} {\bibfnamefont {J.~E.}\ \bibnamefont
  {Bradby}}, \bibinfo {author} {\bibfnamefont {M.~W.}\ \bibnamefont {Doherty}},
  \bibinfo {author} {\bibfnamefont {L.~C.~L.}\ \bibnamefont {Hollenberg}}, \
  and\ \bibinfo {author} {\bibfnamefont {J.-P.}\ \bibnamefont {Tetienne}},\
  }\href@noop {} {\bibfield  {journal} {\bibinfo  {journal} {arXiv:1812.01152}\
  } (\bibinfo {year} {2018})}\BibitemShut {NoStop}%
\bibitem [{\citenamefont {Rajendran}\ \emph {et~al.}(2017)\citenamefont
  {Rajendran}, \citenamefont {Zobrist}, \citenamefont {Sushkov}, \citenamefont
  {Walsworth},\ and\ \citenamefont {Lukin}}]{nvWIMPdetector}%
  \BibitemOpen
  \bibfield  {author} {\bibinfo {author} {\bibfnamefont {S.}~\bibnamefont
  {Rajendran}}, \bibinfo {author} {\bibfnamefont {N.}~\bibnamefont {Zobrist}},
  \bibinfo {author} {\bibfnamefont {A.~O.}\ \bibnamefont {Sushkov}}, \bibinfo
  {author} {\bibfnamefont {R.}~\bibnamefont {Walsworth}}, \ and\ \bibinfo
  {author} {\bibfnamefont {M.}~\bibnamefont {Lukin}},\ }\href {\doibase
  10.1103/PhysRevD.96.035009} {\bibfield  {journal} {\bibinfo  {journal} {Phys.
  Rev. D}\ }\textbf {\bibinfo {volume} {96}},\ \bibinfo {pages} {035009}
  (\bibinfo {year} {2017})}\BibitemShut {NoStop}%
\bibitem [{\citenamefont {Barfuss}\ \emph {et~al.}(2019)\citenamefont
  {Barfuss}, \citenamefont {Kasperczyk}, \citenamefont {K\"olbl},\ and\
  \citenamefont {Maletinsky}}]{maletinskyStrainTerms}%
  \BibitemOpen
  \bibfield  {author} {\bibinfo {author} {\bibfnamefont {A.}~\bibnamefont
  {Barfuss}}, \bibinfo {author} {\bibfnamefont {M.}~\bibnamefont {Kasperczyk}},
  \bibinfo {author} {\bibfnamefont {J.}~\bibnamefont {K\"olbl}}, \ and\
  \bibinfo {author} {\bibfnamefont {P.}~\bibnamefont {Maletinsky}},\ }\href
  {\doibase 10.1103/PhysRevB.99.174102} {\bibfield  {journal} {\bibinfo
  {journal} {Phys. Rev. B}\ }\textbf {\bibinfo {volume} {99}},\ \bibinfo
  {pages} {174102} (\bibinfo {year} {2019})}\BibitemShut {NoStop}%
\bibitem [{\citenamefont {Barson}\ \emph {et~al.}(2017)\citenamefont {Barson},
  \citenamefont {Peddibhotla}, \citenamefont {Ovartchaiyapong}, \citenamefont
  {Ganesan}, \citenamefont {Taylor}, \citenamefont {Gebert}, \citenamefont
  {Mielens}, \citenamefont {Koslowski}, \citenamefont {Simpson}, \citenamefont
  {McGuinness}, \citenamefont {McCallum}, \citenamefont {Prawer}, \citenamefont
  {Onoda}, \citenamefont {Ohshima}, \citenamefont {Bleszynski~Jayich},
  \citenamefont {Jelezko}, \citenamefont {Manson},\ and\ \citenamefont
  {Doherty}}]{marcusStrainHam}%
  \BibitemOpen
  \bibfield  {author} {\bibinfo {author} {\bibfnamefont {M.~S.~J.}\
  \bibnamefont {Barson}}, \bibinfo {author} {\bibfnamefont {P.}~\bibnamefont
  {Peddibhotla}}, \bibinfo {author} {\bibfnamefont {P.}~\bibnamefont
  {Ovartchaiyapong}}, \bibinfo {author} {\bibfnamefont {K.}~\bibnamefont
  {Ganesan}}, \bibinfo {author} {\bibfnamefont {R.~L.}\ \bibnamefont {Taylor}},
  \bibinfo {author} {\bibfnamefont {M.}~\bibnamefont {Gebert}}, \bibinfo
  {author} {\bibfnamefont {Z.}~\bibnamefont {Mielens}}, \bibinfo {author}
  {\bibfnamefont {B.}~\bibnamefont {Koslowski}}, \bibinfo {author}
  {\bibfnamefont {D.~A.}\ \bibnamefont {Simpson}}, \bibinfo {author}
  {\bibfnamefont {L.~P.}\ \bibnamefont {McGuinness}}, \bibinfo {author}
  {\bibfnamefont {J.}~\bibnamefont {McCallum}}, \bibinfo {author}
  {\bibfnamefont {S.}~\bibnamefont {Prawer}}, \bibinfo {author} {\bibfnamefont
  {S.}~\bibnamefont {Onoda}}, \bibinfo {author} {\bibfnamefont
  {T.}~\bibnamefont {Ohshima}}, \bibinfo {author} {\bibfnamefont {A.~C.}\
  \bibnamefont {Bleszynski~Jayich}}, \bibinfo {author} {\bibfnamefont
  {F.}~\bibnamefont {Jelezko}}, \bibinfo {author} {\bibfnamefont {N.~B.}\
  \bibnamefont {Manson}}, \ and\ \bibinfo {author} {\bibfnamefont {M.~W.}\
  \bibnamefont {Doherty}},\ }\href {\doibase 10.1021/acs.nanolett.6b04544}
  {\bibfield  {journal} {\bibinfo  {journal} {Nano Letters}\ }\textbf {\bibinfo
  {volume} {17}},\ \bibinfo {pages} {1496} (\bibinfo {year}
  {2017})}\BibitemShut {NoStop}%
\bibitem [{\citenamefont {Udvarhelyi}\ \emph {et~al.}(2018)\citenamefont
  {Udvarhelyi}, \citenamefont {Shkolnikov}, \citenamefont {Gali}, \citenamefont
  {Burkard},\ and\ \citenamefont {P\'alyi}}]{galiSpinStrain}%
  \BibitemOpen
  \bibfield  {author} {\bibinfo {author} {\bibfnamefont {P.}~\bibnamefont
  {Udvarhelyi}}, \bibinfo {author} {\bibfnamefont {V.~O.}\ \bibnamefont
  {Shkolnikov}}, \bibinfo {author} {\bibfnamefont {A.}~\bibnamefont {Gali}},
  \bibinfo {author} {\bibfnamefont {G.}~\bibnamefont {Burkard}}, \ and\
  \bibinfo {author} {\bibfnamefont {A.}~\bibnamefont {P\'alyi}},\ }\href
  {\doibase 10.1103/PhysRevB.98.075201} {\bibfield  {journal} {\bibinfo
  {journal} {Phys. Rev. B}\ }\textbf {\bibinfo {volume} {98}},\ \bibinfo
  {pages} {075201} (\bibinfo {year} {2018})}\BibitemShut {NoStop}%
\bibitem [{\citenamefont {Acosta}\ \emph {et~al.}(2010)\citenamefont {Acosta},
  \citenamefont {Bauch}, \citenamefont {Ledbetter}, \citenamefont {Waxman},
  \citenamefont {Bouchard},\ and\ \citenamefont {Budker}}]{victor_Tdepend}%
  \BibitemOpen
  \bibfield  {author} {\bibinfo {author} {\bibfnamefont {V.~M.}\ \bibnamefont
  {Acosta}}, \bibinfo {author} {\bibfnamefont {E.}~\bibnamefont {Bauch}},
  \bibinfo {author} {\bibfnamefont {M.~P.}\ \bibnamefont {Ledbetter}}, \bibinfo
  {author} {\bibfnamefont {A.}~\bibnamefont {Waxman}}, \bibinfo {author}
  {\bibfnamefont {L.-S.}\ \bibnamefont {Bouchard}}, \ and\ \bibinfo {author}
  {\bibfnamefont {D.}~\bibnamefont {Budker}},\ }\href {\doibase
  10.1103/PhysRevLett.104.070801} {\bibfield  {journal} {\bibinfo  {journal}
  {Phys. Rev. Lett.}\ }\textbf {\bibinfo {volume} {104}},\ \bibinfo {pages}
  {070801} (\bibinfo {year} {2010})}\BibitemShut {NoStop}%
\bibitem [{\citenamefont {Doherty}\ \emph {et~al.}(2014)\citenamefont
  {Doherty}, \citenamefont {Acosta}, \citenamefont {Jarmola}, \citenamefont
  {Barson}, \citenamefont {Manson}, \citenamefont {Budker},\ and\ \citenamefont
  {Hollenberg}}]{marcus_temperature}%
  \BibitemOpen
  \bibfield  {author} {\bibinfo {author} {\bibfnamefont {M.~W.}\ \bibnamefont
  {Doherty}}, \bibinfo {author} {\bibfnamefont {V.~M.}\ \bibnamefont {Acosta}},
  \bibinfo {author} {\bibfnamefont {A.}~\bibnamefont {Jarmola}}, \bibinfo
  {author} {\bibfnamefont {M.~S.~J.}\ \bibnamefont {Barson}}, \bibinfo {author}
  {\bibfnamefont {N.~B.}\ \bibnamefont {Manson}}, \bibinfo {author}
  {\bibfnamefont {D.}~\bibnamefont {Budker}}, \ and\ \bibinfo {author}
  {\bibfnamefont {L.~C.~L.}\ \bibnamefont {Hollenberg}},\ }\href {\doibase
  10.1103/PhysRevB.90.041201} {\bibfield  {journal} {\bibinfo  {journal} {Phys.
  Rev. B}\ }\textbf {\bibinfo {volume} {90}},\ \bibinfo {pages} {041201}
  (\bibinfo {year} {2014})}\BibitemShut {NoStop}%
\bibitem [{\citenamefont {Hoa}\ \emph {et~al.}(2014)\citenamefont {Hoa},
  \citenamefont {Ouisse}, \citenamefont {Chaussende}, \citenamefont {Naamoun},
  \citenamefont {Tallaire},\ and\ \citenamefont {Achard}}]{hoa2014}%
  \BibitemOpen
  \bibfield  {author} {\bibinfo {author} {\bibfnamefont {L.~T.~M.}\
  \bibnamefont {Hoa}}, \bibinfo {author} {\bibfnamefont {T.}~\bibnamefont
  {Ouisse}}, \bibinfo {author} {\bibfnamefont {D.}~\bibnamefont {Chaussende}},
  \bibinfo {author} {\bibfnamefont {M.}~\bibnamefont {Naamoun}}, \bibinfo
  {author} {\bibfnamefont {A.}~\bibnamefont {Tallaire}}, \ and\ \bibinfo
  {author} {\bibfnamefont {J.}~\bibnamefont {Achard}},\ }\href {\doibase
  10.1021/cg5010193} {\bibfield  {journal} {\bibinfo  {journal} {Crystal Growth
  \& Design}\ }\textbf {\bibinfo {volume} {14}},\ \bibinfo {pages} {5761}
  (\bibinfo {year} {2014})}\BibitemShut {NoStop}%
\bibitem [{\citenamefont {Glazer}\ \emph {et~al.}(1996)\citenamefont {Glazer},
  \citenamefont {Lewis},\ and\ \citenamefont {Kaminsky}}]{glazer1996}%
  \BibitemOpen
  \bibfield  {author} {\bibinfo {author} {\bibfnamefont {A.~M.}\ \bibnamefont
  {Glazer}}, \bibinfo {author} {\bibfnamefont {J.~G.}\ \bibnamefont {Lewis}}, \
  and\ \bibinfo {author} {\bibfnamefont {W.}~\bibnamefont {Kaminsky}},\ }\href
  {\doibase 10.1098/rspa.1996.0145} {\bibfield  {journal} {\bibinfo  {journal}
  {Proceedings of the Royal Society of London A: Mathematical, Physical and
  Engineering Sciences}\ }\textbf {\bibinfo {volume} {452}},\ \bibinfo {pages}
  {2751} (\bibinfo {year} {1996})}\BibitemShut {NoStop}%
\bibitem [{\citenamefont {Kaminsky}\ \emph {et~al.}(2007)\citenamefont
  {Kaminsky}, \citenamefont {Gunn}, \citenamefont {Sours},\ and\ \citenamefont
  {Kahr}}]{kaminsky2007}%
  \BibitemOpen
  \bibfield  {author} {\bibinfo {author} {\bibfnamefont {W.}~\bibnamefont
  {Kaminsky}}, \bibinfo {author} {\bibfnamefont {E.}~\bibnamefont {Gunn}},
  \bibinfo {author} {\bibfnamefont {R.}~\bibnamefont {Sours}}, \ and\ \bibinfo
  {author} {\bibfnamefont {B.}~\bibnamefont {Kahr}},\ }\href {\doibase
  10.1111/j.1365-2818.2007.01841.x} {\bibfield  {journal} {\bibinfo  {journal}
  {Journal of Microscopy}\ }\textbf {\bibinfo {volume} {228}},\ \bibinfo
  {pages} {153} (\bibinfo {year} {2007})}\BibitemShut {NoStop}%
\bibitem [{\citenamefont {Howell}(2012)}]{howell2012Review}%
  \BibitemOpen
  \bibfield  {author} {\bibinfo {author} {\bibfnamefont {D.}~\bibnamefont
  {Howell}},\ }\href {\doibase 10.1127/0935-1221/2012/0024-2205} {\bibfield
  {journal} {\bibinfo  {journal} {European Journal of Mineralogy}\ }\textbf
  {\bibinfo {volume} {24}},\ \bibinfo {pages} {575} (\bibinfo {year}
  {2012})}\BibitemShut {NoStop}%
\bibitem [{\citenamefont {Nye}(1957)}]{nye1957review}%
  \BibitemOpen
  \bibfield  {author} {\bibinfo {author} {\bibfnamefont {J.~F.}\ \bibnamefont
  {Nye}},\ }\href@noop {} {\emph {\bibinfo {title} {Physical Properties of
  Crystals}}}\ (\bibinfo  {publisher} {Oxford University Press},\ \bibinfo
  {year} {1957})\BibitemShut {NoStop}%
\bibitem [{\citenamefont {Ramachandran}(1947)}]{Ramachandran1947}%
  \BibitemOpen
  \bibfield  {author} {\bibinfo {author} {\bibfnamefont {G.~N.}\ \bibnamefont
  {Ramachandran}},\ }\href {\doibase 10.1007/BF03170953} {\bibfield  {journal}
  {\bibinfo  {journal} {Proceedings of the Indian Academy of Sciences - Section
  A}\ }\textbf {\bibinfo {volume} {26}},\ \bibinfo {pages} {77} (\bibinfo
  {year} {1947})}\BibitemShut {NoStop}%
\bibitem [{\citenamefont {Hounsome}\ \emph {et~al.}(2006)\citenamefont
  {Hounsome}, \citenamefont {Jones}, \citenamefont {Shaw},\ and\ \citenamefont
  {Briddon}}]{hounsome2006}%
  \BibitemOpen
  \bibfield  {author} {\bibinfo {author} {\bibfnamefont {L.~S.}\ \bibnamefont
  {Hounsome}}, \bibinfo {author} {\bibfnamefont {R.}~\bibnamefont {Jones}},
  \bibinfo {author} {\bibfnamefont {M.~J.}\ \bibnamefont {Shaw}}, \ and\
  \bibinfo {author} {\bibfnamefont {P.~R.}\ \bibnamefont {Briddon}},\ }\href
  {\doibase 10.1002/pssa.200671121} {\bibfield  {journal} {\bibinfo  {journal}
  {physica status solidi (a)}\ }\textbf {\bibinfo {volume} {203}},\ \bibinfo
  {pages} {3088} (\bibinfo {year} {2006})}\BibitemShut {NoStop}%
\bibitem [{\citenamefont {Jensen}\ \emph {et~al.}(2017)\citenamefont {Jensen},
  \citenamefont {Kehayias},\ and\ \citenamefont {Budker}}]{kasperBookCh}%
  \BibitemOpen
  \bibfield  {author} {\bibinfo {author} {\bibfnamefont {K.}~\bibnamefont
  {Jensen}}, \bibinfo {author} {\bibfnamefont {P.}~\bibnamefont {Kehayias}}, \
  and\ \bibinfo {author} {\bibfnamefont {D.}~\bibnamefont {Budker}},\ }\enquote
  {\bibinfo {title} {Magnetometry with nitrogen-vacancy centers in diamond},}\
  in\ \href {\doibase 10.1007/978-3-319-34070-8_18} {\emph {\bibinfo
  {booktitle} {High Sensitivity Magnetometers}}},\ \bibinfo {editor} {edited
  by\ \bibinfo {editor} {\bibfnamefont {A.}~\bibnamefont {Grosz}}, \bibinfo
  {editor} {\bibfnamefont {M.~J.}\ \bibnamefont {Haji-Sheikh}}, \ and\ \bibinfo
  {editor} {\bibfnamefont {S.~C.}\ \bibnamefont {Mukhopadhyay}}}\ (\bibinfo
  {publisher} {Springer International Publishing},\ \bibinfo {address} {Cham},\
  \bibinfo {year} {2017})\ pp.\ \bibinfo {pages} {553--576}\BibitemShut
  {NoStop}%
\bibitem [{\citenamefont {Glenn}\ \emph {et~al.}(2018)\citenamefont {Glenn},
  \citenamefont {Bucher}, \citenamefont {Lee}, \citenamefont {Lukin},
  \citenamefont {Park},\ and\ \citenamefont {Walsworth}}]{Glenn2018}%
  \BibitemOpen
  \bibfield  {author} {\bibinfo {author} {\bibfnamefont {D.~R.}\ \bibnamefont
  {Glenn}}, \bibinfo {author} {\bibfnamefont {D.~B.}\ \bibnamefont {Bucher}},
  \bibinfo {author} {\bibfnamefont {J.}~\bibnamefont {Lee}}, \bibinfo {author}
  {\bibfnamefont {M.~D.}\ \bibnamefont {Lukin}}, \bibinfo {author}
  {\bibfnamefont {H.}~\bibnamefont {Park}}, \ and\ \bibinfo {author}
  {\bibfnamefont {R.~L.}\ \bibnamefont {Walsworth}},\ }\href
  {http://dx.doi.org/10.1038/nature25781} {\bibfield  {journal} {\bibinfo
  {journal} {Nature}\ }\textbf {\bibinfo {volume} {555}},\ \bibinfo {pages}
  {351 EP } (\bibinfo {year} {2018})}\BibitemShut {NoStop}%
\bibitem [{\citenamefont {Gaukroger}\ \emph {et~al.}(2008)\citenamefont
  {Gaukroger}, \citenamefont {Martineau}, \citenamefont {Crowder},
  \citenamefont {Friel}, \citenamefont {Williams},\ and\ \citenamefont
  {Twitchen}}]{twitchenXray}%
  \BibitemOpen
  \bibfield  {author} {\bibinfo {author} {\bibfnamefont {M.}~\bibnamefont
  {Gaukroger}}, \bibinfo {author} {\bibfnamefont {P.}~\bibnamefont
  {Martineau}}, \bibinfo {author} {\bibfnamefont {M.}~\bibnamefont {Crowder}},
  \bibinfo {author} {\bibfnamefont {I.}~\bibnamefont {Friel}}, \bibinfo
  {author} {\bibfnamefont {S.}~\bibnamefont {Williams}}, \ and\ \bibinfo
  {author} {\bibfnamefont {D.}~\bibnamefont {Twitchen}},\ }\href {\doibase
  https://doi.org/10.1016/j.diamond.2007.12.036} {\bibfield  {journal}
  {\bibinfo  {journal} {Diamond and Related Materials}\ }\textbf {\bibinfo
  {volume} {17}},\ \bibinfo {pages} {262 } (\bibinfo {year}
  {2008})}\BibitemShut {NoStop}%
\bibitem [{\citenamefont {Tsubouchi}\ \emph {et~al.}(2009)\citenamefont
  {Tsubouchi}, \citenamefont {Mokuno}, \citenamefont {Yamaguchi}, \citenamefont
  {Tatsumi}, \citenamefont {Chayahara},\ and\ \citenamefont
  {Shikata}}]{shikataXray}%
  \BibitemOpen
  \bibfield  {author} {\bibinfo {author} {\bibfnamefont {N.}~\bibnamefont
  {Tsubouchi}}, \bibinfo {author} {\bibfnamefont {Y.}~\bibnamefont {Mokuno}},
  \bibinfo {author} {\bibfnamefont {H.}~\bibnamefont {Yamaguchi}}, \bibinfo
  {author} {\bibfnamefont {N.}~\bibnamefont {Tatsumi}}, \bibinfo {author}
  {\bibfnamefont {A.}~\bibnamefont {Chayahara}}, \ and\ \bibinfo {author}
  {\bibfnamefont {S.}~\bibnamefont {Shikata}},\ }\href {\doibase
  https://doi.org/10.1016/j.diamond.2008.07.022} {\bibfield  {journal}
  {\bibinfo  {journal} {Diamond and Related Materials}\ }\textbf {\bibinfo
  {volume} {18}},\ \bibinfo {pages} {216 } (\bibinfo {year} {2009})},\ \bibinfo
  {note} {proceedings of the International Conference on New Diamond and Nano
  Carbons 2008}\BibitemShut {NoStop}%
\bibitem [{\citenamefont {Martineau}\ \emph {et~al.}(2004)\citenamefont
  {Martineau}, \citenamefont {Lawson}, \citenamefont {Taylor}, \citenamefont
  {Quinn}, \citenamefont {Evans},\ and\ \citenamefont
  {Crowder}}]{martineauXray}%
  \BibitemOpen
  \bibfield  {author} {\bibinfo {author} {\bibfnamefont {P.~M.}\ \bibnamefont
  {Martineau}}, \bibinfo {author} {\bibfnamefont {S.~C.}\ \bibnamefont
  {Lawson}}, \bibinfo {author} {\bibfnamefont {A.~J.}\ \bibnamefont {Taylor}},
  \bibinfo {author} {\bibfnamefont {S.~J.}\ \bibnamefont {Quinn}}, \bibinfo
  {author} {\bibfnamefont {D.~J.~F.}\ \bibnamefont {Evans}}, \ and\ \bibinfo
  {author} {\bibfnamefont {M.~J.}\ \bibnamefont {Crowder}},\ }\href@noop {}
  {\bibfield  {journal} {\bibinfo  {journal} {Gems \& Gemology}\ }\textbf
  {\bibinfo {volume} {40}},\ \bibinfo {pages} {2} (\bibinfo {year}
  {2004})}\BibitemShut {NoStop}%
\bibitem [{\citenamefont {Hull}\ and\ \citenamefont {Bacon}(2011)}]{hullBacon}%
  \BibitemOpen
  \bibfield  {author} {\bibinfo {author} {\bibfnamefont {D.}~\bibnamefont
  {Hull}}\ and\ \bibinfo {author} {\bibfnamefont {D.}~\bibnamefont {Bacon}},\
  }\href {\doibase https://doi.org/10.1016/B978-0-08-096672-4.00004-9} {\emph
  {\bibinfo {title} {Introduction to Dislocations (Fifth Edition)}}}\ (\bibinfo
   {publisher} {Butterworth-Heinemann},\ \bibinfo {address} {Oxford},\ \bibinfo
  {year} {2011})\BibitemShut {NoStop}%
\bibitem [{\citenamefont {Pinto}\ and\ \citenamefont
  {Jones}(2009)}]{Pinto2009}%
  \BibitemOpen
  \bibfield  {author} {\bibinfo {author} {\bibfnamefont {H.}~\bibnamefont
  {Pinto}}\ and\ \bibinfo {author} {\bibfnamefont {R.}~\bibnamefont {Jones}},\
  }\href {http://stacks.iop.org/0953-8984/21/i=36/a=364220} {\bibfield
  {journal} {\bibinfo  {journal} {Journal of Physics: Condensed Matter}\
  }\textbf {\bibinfo {volume} {21}},\ \bibinfo {pages} {364220} (\bibinfo
  {year} {2009})}\BibitemShut {NoStop}%
\bibitem [{\citenamefont {Toyli}\ \emph {et~al.}(2013)\citenamefont {Toyli},
  \citenamefont {de~las Casas}, \citenamefont {Christle}, \citenamefont
  {Dobrovitski},\ and\ \citenamefont {Awschalom}}]{ToyliTemperature}%
  \BibitemOpen
  \bibfield  {author} {\bibinfo {author} {\bibfnamefont {D.~M.}\ \bibnamefont
  {Toyli}}, \bibinfo {author} {\bibfnamefont {C.~F.}\ \bibnamefont {de~las
  Casas}}, \bibinfo {author} {\bibfnamefont {D.~J.}\ \bibnamefont {Christle}},
  \bibinfo {author} {\bibfnamefont {V.~V.}\ \bibnamefont {Dobrovitski}}, \ and\
  \bibinfo {author} {\bibfnamefont {D.~D.}\ \bibnamefont {Awschalom}},\ }\href
  {\doibase 10.1073/pnas.1306825110} {\bibfield  {journal} {\bibinfo  {journal}
  {Proceedings of the National Academy of Sciences}\ }\textbf {\bibinfo
  {volume} {110}},\ \bibinfo {pages} {8417} (\bibinfo {year}
  {2013})}\BibitemShut {NoStop}%
\bibitem [{\citenamefont {Rittweger}\ \emph {et~al.}(2009)\citenamefont
  {Rittweger}, \citenamefont {Han}, \citenamefont {Irvine}, \citenamefont
  {Eggeling},\ and\ \citenamefont {Hell}}]{STED1}%
  \BibitemOpen
  \bibfield  {author} {\bibinfo {author} {\bibfnamefont {E.}~\bibnamefont
  {Rittweger}}, \bibinfo {author} {\bibfnamefont {K.~Y.}\ \bibnamefont {Han}},
  \bibinfo {author} {\bibfnamefont {S.~E.}\ \bibnamefont {Irvine}}, \bibinfo
  {author} {\bibfnamefont {C.}~\bibnamefont {Eggeling}}, \ and\ \bibinfo
  {author} {\bibfnamefont {S.~W.}\ \bibnamefont {Hell}},\ }\href@noop {}
  {\bibfield  {journal} {\bibinfo  {journal} {Nat Photon}\ }\textbf {\bibinfo
  {volume} {3}},\ \bibinfo {pages} {144} (\bibinfo {year} {2009})}\BibitemShut
  {NoStop}%
\bibitem [{\citenamefont {Jaskula}\ \emph {et~al.}(2017)\citenamefont
  {Jaskula}, \citenamefont {Bauch}, \citenamefont {Arroyo-Camejo},
  \citenamefont {Lukin}, \citenamefont {Hell}, \citenamefont {Trifonov},\ and\
  \citenamefont {Walsworth}}]{JCsuperres}%
  \BibitemOpen
  \bibfield  {author} {\bibinfo {author} {\bibfnamefont {J.-C.}\ \bibnamefont
  {Jaskula}}, \bibinfo {author} {\bibfnamefont {E.}~\bibnamefont {Bauch}},
  \bibinfo {author} {\bibfnamefont {S.}~\bibnamefont {Arroyo-Camejo}}, \bibinfo
  {author} {\bibfnamefont {M.~D.}\ \bibnamefont {Lukin}}, \bibinfo {author}
  {\bibfnamefont {S.~W.}\ \bibnamefont {Hell}}, \bibinfo {author}
  {\bibfnamefont {A.~S.}\ \bibnamefont {Trifonov}}, \ and\ \bibinfo {author}
  {\bibfnamefont {R.~L.}\ \bibnamefont {Walsworth}},\ }\href {\doibase
  10.1364/OE.25.011048} {\bibfield  {journal} {\bibinfo  {journal} {Opt.
  Express}\ }\textbf {\bibinfo {volume} {25}},\ \bibinfo {pages} {11048}
  (\bibinfo {year} {2017})}\BibitemShut {NoStop}%
\end{thebibliography}

%

\newpage



\section*{Supplementary material}

# Supplemental material for "Imaging crystal stress in diamond using ensembles of nitrogen-vacancy centers"

P. Kehayias,[1, 2, *] M. J. Turner,[1, 3] R. Trubko,[1] J. M. Schloss,[3, 4]
C. A. Hart,[1] M. Wesson,[5] D. R. Glenn,[1] and R. L. Walsworth[1, 2, 3]

[1]*Department of Physics, Harvard University, Cambridge, MA 02138, USA*
[2]*Harvard-Smithsonian Center for Astrophysics, Cambridge, MA 02138, USA*
[3]*Center for Brain Science, Harvard University, Cambridge, MA 02138, USA*
[4]*Department of Physics, Massachusetts Institute of Technology, Cambridge, MA 02139, USA*
[5]*Department of Physics, University of Chicago, Chicago, IL 60637, USA*

## CONTENTS

## FULL DETAILS OF DIAMOND SAMPLES INVESTIGATED

### Selected regions of Sample A

Figure S1 shows the regions on Sample A selected for further analysis in other figures. This displays the fields-of-view used to illustrate how different strain features affect NV magnetometry and how NV stress mapping compares to birefringence imaging.

### Summary of additional diamond samples studied

Table S1 lists the properties for nine additional diamond samples we studied (labeled Sample B through Sample J), with NV layer thicknesses ranging from 20 nm to 140 μm. We include these samples to provide additional examples of various strain features. NV $M_{z,\kappa}$ and stress maps for the additional samples are included in the last section.

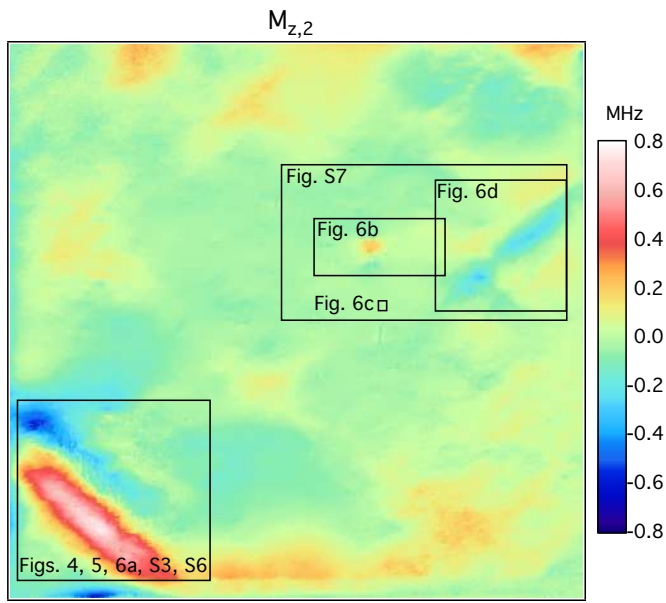

**Supplemental Figure** S1. Field-of-view locations on Sample A used for other figures in the main text and supplemental material.

| Sample | Dimensions | [N] in layer | Layer Thickness | Substrate | Irradiated/ Annealed? | References |
|---|---|---|---|---|---|---|
| A | 4×4×0.5 mm³ | 25 ppm | 13 μm | CVD | Yes / Yes | Studied in the main text |
| B | 4×4×0.5 mm³ | 20 ppm | 4 μm | CVD | Yes / Yes | Sample D4 in Ref. [S1] |
| C | 5×5×0.4 mm³ | 10 ppm | 40 μm | CVD | Yes / Yes | - |
| D | 2×2×0.5 mm³ | 2×10¹¹/cm² dose | 20 nm | CVD | 14 keV ¹⁴N⁺ / Yes | Sample D1 in Ref. [S1] |
| E | 1.7×1.5×0.5 mm³ | 10 ppm | 40 μm | CVD | No / No | Sample C in Ref. [S2] |
| F | 4×4×0.3 mm³ | 0.75 ppm | 140 μm | HPHT | No / No | Sample B in Ref. [S2] |
| G | 4.5×4.5×0.5 mm³ | 20 ppm | 4 μm | CVD | Yes / Yes | Used in Ref. [S3] |
| H | 4×4×0.5 mm³ | 27 ppm | 13 μm | CVD | Yes / Yes | - |
| I | 4×4×0.5 mm³ | 7.2 ppm | 0.9 μm | CVD | Yes / Yes | Sample D2 in Ref. [S4] |
| J | 4×4×0.5 mm³ | 26.8 ppm | 13 μm | CVD | Yes / Yes | Sample D3 in Ref. [S1] |

**Supplemental Table** S1. Properties of all diamond samples studied in this work. Sample F has a nitrogen-enriched chemical vapor deposition (CVD) layer grown on top of a diamond substrate made by high-pressure high-temperature (HPHT) synthesis. All other samples have electronic-grade single-crystal (ELSC) substrates grown by CVD. Sample D is a ¹⁴N⁺ beam implant, and the other samples were grown with a nitrogen-rich layer.

## SPIN-STRESS TERMS IN THE NV HAMILTONIAN

### Corrections from neglected spin-stress terms

Using second-order perturbation theory, we estimate the potential corrections of the neglected spin-stress terms $\{M_{x,\kappa}, M_{y,\kappa}, N_{x,\kappa}, N_{y,\kappa}\}$ on the NV transition frequencies. These calculations justify disregarding these terms in our analysis, keeping only the $M_{z,\kappa}$ spin-stress term. For simplicity, we do this analysis for the projection magnetic microscopy case, where $\vec{B} = B_z\hat{z}$, with $B_z \approx 1$ mT in our experiments. Note that the contributions from the neglected spin-stress terms may become significant if the Zeeman effect is small in comparison, which can happen if $\vec{B}$ is largely perpendicular to the NV $z$-axis in a vector magnetic microscopy experiment.

For the $M_{x,\kappa}$ and $M_{y,\kappa}$ spin-stress terms, the $m_s = \pm 1$ electronic sublevel energies are shifted by $\pm \frac{M_{x,\kappa}^2 + M_{y,\kappa}^2}{2\gamma B_z}$ while the $m_s = 0$ electronic sublevel is unaffected. For a typical 100 kHz spin-stress contribution, this perturbation shifts the NV transition frequencies by 200 Hz, which is a $6\times10^{-8}$ fractional change.

For the $N_{x,\kappa}$ and $N_{y,\kappa}$ spin-stress terms, the $m_s = \pm 1$ electronic sublevel energies are shifted by $\frac{\pm(N_{x,\kappa}^2 + N_{y,\kappa}^2)}{2(D + M_{z,\kappa} \pm \gamma B_z)}$ while the $m_s = 0$ electronic sublevel is shifted by $-\frac{N_{x,\kappa}^2 + N_{y,\kappa}^2}{2(D + M_{z,\kappa} + \gamma B_z)} - \frac{N_{x,\kappa}^2 + N_{y,\kappa}^2}{2(D + M_{z,\kappa} - \gamma B_z)}$. When $\gamma B_z \ll D$ and $M_{z,\kappa} \ll D$, the NV transition frequencies shift by approximately $\frac{3(N_{x,\kappa}^2 + N_{y,\kappa}^2)}{2D}$ and $\frac{N_{x,\kappa}^2 + N_{y,\kappa}^2}{2D}$. For a 100 kHz spin-stress contribution, this perturbation shifts the NV transition frequencies by at most 5 Hz, which is a $2 \times 10^{-9}$ fractional change. Note that this estimate is just for illustration, as experimental values for the $d$ and $e$ spin-stress coupling constants for $N_{x,\kappa}$ and $N_{y,\kappa}$ are not currently reported to our knowledge [S5, S6], though Ref. [S7] calculated estimates for them numerically. The above transition frequency correction estimates justify our rationale to neglect these spin-stress terms in our analysis.

### Stress and electric-field contributions to the NV ground-state Hamiltonian

Crystal stress and electric field are often intertwined through the piezoelectric, pyroelectric, and ferroelectric effects. However, the bonds between carbon atoms in the diamond lattice are covalent bonds, meaning diamond should have no permanent or induced electric dipole moment in its unit cell. Thus, in the absence of defects, stress within diamond does not cause an electric field or vice versa, but this does not mean they are unrelated. For example, an NV center can sense an electric field from nearby charged defects, which can also cause normal stress due to lattice deformation.

From Ref. [S7], the electric-field contributions to the NV ground-state Hamiltonian are

$$
\begin{aligned}
H_{E,\kappa} = {} & d_{||} E_z S_{z,\kappa}^2 + d_\perp [(S_{y,\kappa}^2 - S_{x,\kappa}^2) E_x + (S_{x,\kappa} S_{y,\kappa} + S_{y,\kappa} S_{x,\kappa}) E_y] \\
& + d'_\perp [(S_{x,\kappa} S_{z,\kappa} + S_{z,\kappa} S_{x,\kappa}) E_x + (S_{y,\kappa} S_{z,\kappa} + S_{z,\kappa} S_{y,\kappa}) E_y].
\end{aligned}
\tag{S1}
$$

In this expression, $\vec{E}$ is the electric field in the NV coordinate system, and $\{d_{||}, d_\perp, d'_\perp\}$ are NV ground-state electric dipole parameters [S8]. The spin-stress terms in Eq. 1 in the main text and the $\vec{E}$ terms affect the same spin terms in the NV Hamiltonian, though the former originate from a stress tensor and the latter originate from a vector electric field. When performing vector magnetic microscopy or projection magnetic microscopy with $|\vec{B}| > 1$ mT, only the $M_{z,\kappa}$ and $E_z$ terms matter, and the off-diagonal terms can be ignored.

Because of the similarity between how $M_{z,\kappa}$ and $E_z$ enter the NV Hamiltonian, at first glance it may be unclear whether our NV ODMR measurements are imaging inhomogeneity caused by $M_{z,\kappa}$, $E_z$, or both. Herein we argue that our imaging experiments are primarily sensitive to $M_{z,\kappa}$, while $E_z$ may contribute to lineshape broadening.

Our diamond samples contain NV ensembles with equal populations oriented along opposite $z$ directions (for example, [111] and [$\bar{1}\bar{1}1$]). There are eight NV $z$ directions (two for each $\kappa$). Usually the NVs pointing along opposite directions behave identically, allowing us to group them together with the same $\kappa$ label. Calculating the stress tensor and the $M_{z,\kappa}$ for each NV orientation (Tab. S2) we see that the NV sub-ensembles with $z$-axes pointing in opposite directions have the same $M_{z,\kappa}$ and the same resonance frequencies.

Although the stress tensor contributions affect the [111] and [$\bar{1}\bar{1}1$] NV orientations the same way, this is not true for the $E_z$ electric field contribution. The $E_z$ term shifts the NV resonance frequencies by equal and opposite amounts ($\pm d_{||} E_z$) for NVs with opposite $z$-axes. This causes a lineshape splitting or broadening rather than a common-mode shift. We did not observe such splittings in our NV magnetic resonance spectra (Fig. 2b in the main text), though a weak $E_z$ may contribute as a linewidth broadening. Therefore, the common-mode NV resonance frequency shifts we measure are caused by $M_{z,\kappa}$ terms only. One can imagine extracting $E_z$ from the NV resonance linewidths, but this may be difficult because the linewidth also depends on the magnetic field inhomogeneity, $M_{z,\kappa}$ inhomogeneity, microwave power, and laser power.

### $M_{z,\kappa}$ imaging using NV vector magnetic microscopy and projection magnetic microscopy

As described in the main text, both the vector magnetic microscopy (VMM) and the projection magnetic microscopy (PMM) methods can be used to map $M_{z,\kappa}$ in a diamond [S1]. Figure S2 confirms that the VMM and PMM methods yield the same $M_{z,\kappa}$ information despite the differences between these schemes. VMM yields all $M_{z,\kappa}$ maps simultaneously without having to adjust the bias magnetic field and laser polarization angle (or rotate the diamond). PMM is optimized for measuring one $M_{z,\kappa}$ at a time, and the procedures for lineshape fitting and $M_{z,\kappa}$ extraction are simpler [S1]. As a further check, we rotated the diamond chip by 90° about the $Z$-axis. The resulting $M_{z,\kappa}$ and stress maps were consistent with the unrotated maps, transforming as expected for a 90° rotation.

| $\kappa$ | NV $z$-axes | Stress tensors | | $M_{z,\kappa}$ |
|---|---|---|---|---|
| 1 | $[111]$, $[\bar{1}\bar{1}\bar{1}]$ | $\begin{pmatrix} \sigma_{XX} & \sigma_{XY} & \sigma_{XZ} \\ \sigma_{XY} & \sigma_{YY} & \sigma_{YZ} \\ \sigma_{XZ} & \sigma_{YZ} & \sigma_{ZZ} \end{pmatrix}$ | $\begin{pmatrix} \sigma_{YY} & \sigma_{XY} & \sigma_{YZ} \\ \sigma_{XY} & \sigma_{XX} & \sigma_{XZ} \\ \sigma_{YZ} & \sigma_{XZ} & \sigma_{ZZ} \end{pmatrix}$ | $a_1[\sigma_{XX}+\sigma_{YY}+\sigma_{ZZ}]$ $+2a_2[\sigma_{XY}+\sigma_{XZ}+\sigma_{YZ}]$ |
| 2 | $[\bar{1}\bar{1}1]$, $[11\bar{1}]$ | $\begin{pmatrix} \sigma_{XX} & \sigma_{XY} & -\sigma_{XZ} \\ \sigma_{XY} & \sigma_{YY} & -\sigma_{YZ} \\ -\sigma_{XZ} & -\sigma_{YZ} & \sigma_{ZZ} \end{pmatrix}$ | $\begin{pmatrix} \sigma_{YY} & \sigma_{XY} & -\sigma_{YZ} \\ \sigma_{XY} & \sigma_{XX} & -\sigma_{XZ} \\ -\sigma_{YZ} & -\sigma_{XZ} & \sigma_{ZZ} \end{pmatrix}$ | $a_1[\sigma_{XX}+\sigma_{YY}+\sigma_{ZZ}]$ $+2a_2[\sigma_{XY}-\sigma_{XZ}-\sigma_{YZ}]$ |
| 3 | $[\bar{1}1\bar{1}]$, $[1\bar{1}1]$ | $\begin{pmatrix} \sigma_{XX} & -\sigma_{XY} & \sigma_{XZ} \\ -\sigma_{XY} & \sigma_{YY} & -\sigma_{YZ} \\ \sigma_{XZ} & -\sigma_{YZ} & \sigma_{ZZ} \end{pmatrix}$ | $\begin{pmatrix} \sigma_{YY} & -\sigma_{XY} & -\sigma_{YZ} \\ -\sigma_{XY} & \sigma_{XX} & \sigma_{XZ} \\ -\sigma_{YZ} & \sigma_{XZ} & \sigma_{ZZ} \end{pmatrix}$ | $a_1[\sigma_{XX}+\sigma_{YY}+\sigma_{ZZ}]$ $+2a_2[-\sigma_{XY}+\sigma_{XZ}-\sigma_{YZ}]$ |
| 4 | $[1\bar{1}\bar{1}]$, $[\bar{1}11]$ | $\begin{pmatrix} \sigma_{XX} & -\sigma_{XY} & -\sigma_{XZ} \\ -\sigma_{XY} & \sigma_{YY} & \sigma_{YZ} \\ -\sigma_{XZ} & \sigma_{YZ} & \sigma_{ZZ} \end{pmatrix}$ | $\begin{pmatrix} \sigma_{YY} & -\sigma_{XY} & \sigma_{YZ} \\ -\sigma_{XY} & \sigma_{XX} & -\sigma_{XZ} \\ \sigma_{YZ} & -\sigma_{XZ} & \sigma_{ZZ} \end{pmatrix}$ | $a_1[\sigma_{XX}+\sigma_{YY}+\sigma_{ZZ}]$ $+2a_2[-\sigma_{XY}-\sigma_{XZ}+\sigma_{YZ}]$ |

**Supplemental Table** S2. Stress tensors and $M_{z,\kappa}$ terms for each NV orientation, calculated using the coordinate systems used in Fig. 2a of the main text, and also in Tab. 1 of Ref [S6].

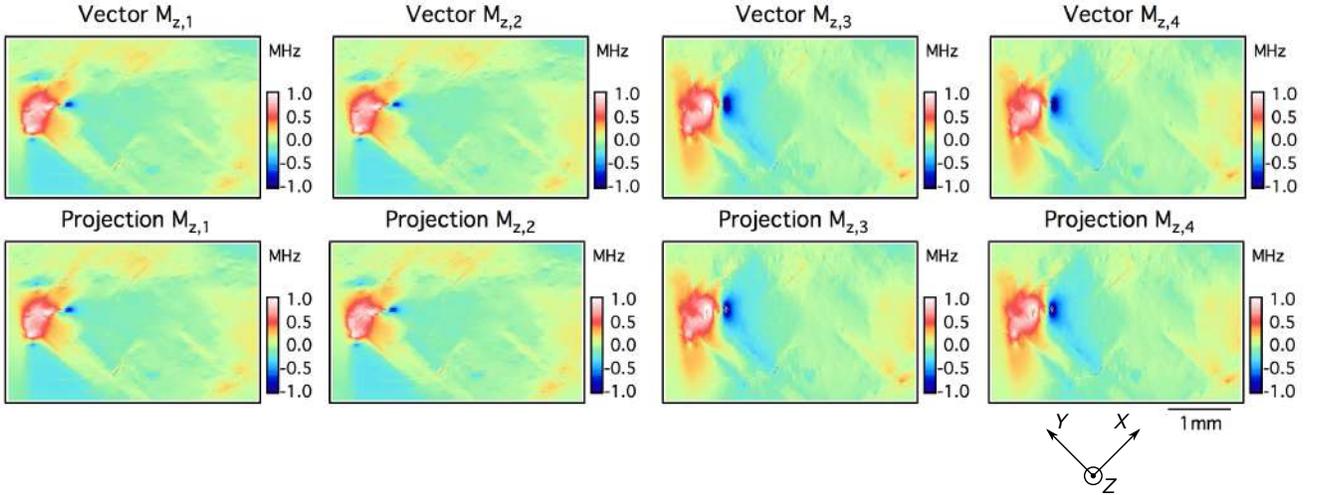

**Supplemental Figure** S2. A comparison of $M_{z,\kappa}$ maps for X-shaped defects in Sample B obtained using vector magnetic microscopy (VMM) and projection magnetic microscopy (PMM). We varied the laser polarization angle and the bias magnetic field in between measurements, keeping the diamond region constant.

### Alternative NV spin-stress coupling constants

We recently became aware of another set of experimental spin-stress coupling constants $\{a_1, a_2\} = \{11.7, -6.5\}$ MHz/GPa reported in Ref. [S6], which differ from the $\{a_1, a_2\} = \{4.86, -3.7\}$ MHz/GPa in Ref. [S5] by roughly a factor of 2. Recalculating the stress maps for our diamond samples using the values of Ref. [S6] produced visually similar stress maps. Furthermore, the calculation in Ref. [S7] reports a third set of $\{a_1, a_2\} = \{2.66, -2.51\}$ MHz/GPa. We do not include stress maps derived from these alternate $\{a_1, a_2\}$ values, though these can be provided upon request. Note that we adopt the convention used in Ref. [S5], where a positive $\sigma_{ii}$ adds a positive contribution to $M_{z,\kappa}$. While we are unable to resolve the disagreement in spin-stress coupling constants, the qualitative messages of our work are unaffected.

### BIREFRINGENCE IMAGING METHODOLOGY AND INTERPRETATION

#### Birefringence imager setup

As illustrated in Fig. 1 in the main text, our NV $M_{z,\kappa}$ imager also includes birefringence imaging capabilities. For birefringence measurements, light from a white LED illuminator passes through a 532 nm bandpass filter and a linear polarizer mounted to a motorized rotation stage before illuminating the diamond sample. The light transmitted

through the birefringent diamond is collected by a 4× or 10× microscope objective. The light then passes through a circular analyzer consisting of a zero-order $\lambda/4$ wave plate aligned at 45° to a linear polarizer. The transmitted light is imaged with a CMOS camera through an eyepiece lens. We save camera images while sweeping the polarizer rotation angle $\alpha_i$ through 180°, usually in 10° steps. To alternate between birefringence and NV $M_{z,\kappa}$ imaging, we swap between the circular analyzer and the long-pass filter before the camera.

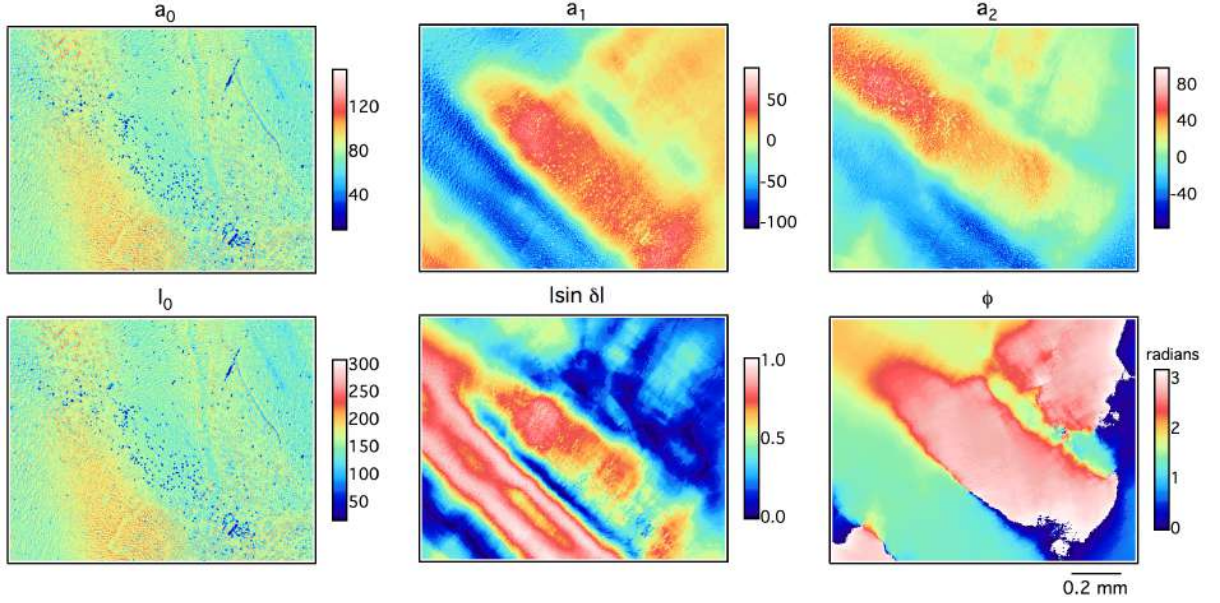

**Supplemental Figure** S3. Complete birefringence maps for Sample A.

### Extracting birefringence parameters

Here we summarize the equations describing rotating-polarizer birefringence measurements, derived in Refs. [S9, S10]. The transmission intensity $I_i$ for polarizer angle $\alpha_i$ is

$$I_i = \frac{1}{2}I_0[1 + \sin 2(\alpha_i - \phi)\sin\delta], \tag{S2}$$

where $I_0$ and $I_i$ are the initial and transmitted intensities, $\phi$ is the retardance orientation angle, $\delta = \frac{2\pi}{\lambda}\Delta nL$ is the optical retardance phase, and $\alpha_i$ is the angle of the first polarizer. This equation can be rewritten in a more convenient form

$$I_i = a_0 + a_1 \sin 2\alpha_i + a_2 \cos 2\alpha_i, \tag{S3}$$

$$a_0 = \frac{1}{2}I_0, \tag{S4}$$

$$a_1 = \frac{1}{2}I_0 \sin\delta\cos 2\phi, \tag{S5}$$

$$a_2 = -\frac{1}{2}I_0 \sin\delta\sin 2\phi. \tag{S6}$$

Solving for the optical parameters of interest yields

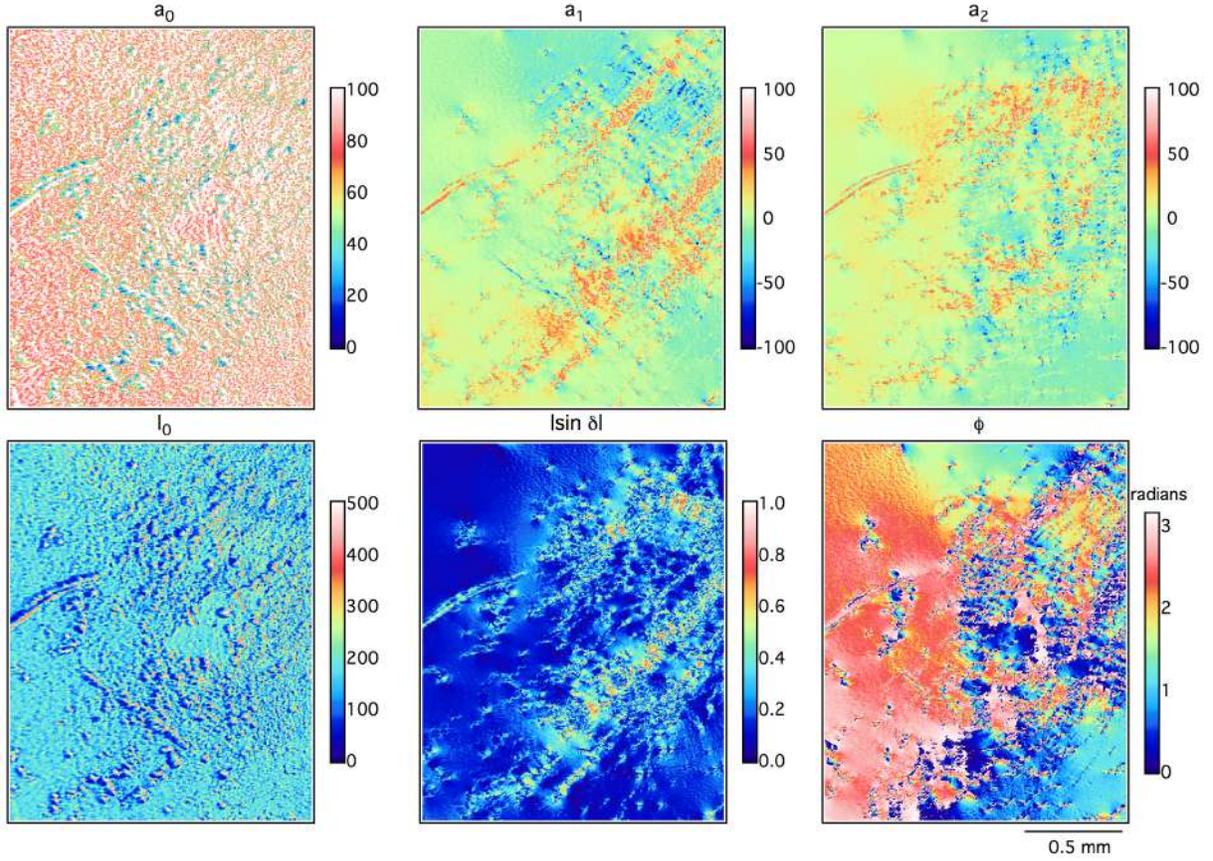

**Supplemental Figure** S4. Complete birefringence maps for Sample C.

$$I_0 = 2a_0, \tag{S7}$$

$$|\sin \delta| = \frac{(a_1{}^2 + a_2{}^2)^{\frac{1}{2}}}{a_0}, \tag{S8}$$

$$\phi = \frac{\pi}{2} + \frac{1}{2}\mathrm{sgn}(a_2) \arccos\left[\frac{-a_1}{(a_1{}^2 + a_2{}^2)^{\frac{1}{2}}}\right]. \tag{S9}$$

One way to extract $a_0$, $a_1$, and $a_2$ from a series of images with varying $\alpha_i$ is to fit intensities in each pixel with Eq. S3. However, this strategy is computationally intensive, especially when imaging over many pixels in a large field of view. Stepping $\alpha_i$ through $N$ equal angles, we can exploit trigonometric properties to write the expressions below as a sum over the angles from $\alpha_1 = 180°/N$ to $\alpha_{\max} = 180°$:

$$a_0 = \frac{1}{N}\sum_{i=1}^{N} I_i, \tag{S10}$$

$$a_1 = \frac{2}{N}\sum_{i=1}^{N} I_i \sin 2\alpha_i, \tag{S11}$$

$$a_2 = \frac{2}{N}\sum_{i=1}^{N} I_i \cos 2\alpha_i, \tag{S12}$$

$$\alpha_1 = 180°/N, \quad \alpha_{\max} = 180°. \tag{S13}$$

Figures S3 and S4 include examples of this reconstruction procedure for $N = 18$ with Samples A and C.

## Birefringence-to-stress approximate model

In order to roughly test the agreement between birefringence imaging and NV stress imaging, we use the photo-elastic equation to convert $\Delta n$ to a stress (or strain) value. The stress and strain inhomogeneity through the diamond is linearly proportional to the $\Delta n$ we measure [S11–S13]. Following these references, we assume an isotropic stress-optic coefficient, $q_{iso}$, and relate stress to $\Delta n$ via

$$|\Delta n| \sim \frac{3}{4} n^3 q_{iso} \sigma \text{ and } |\Delta n| = \frac{|\delta|\lambda}{2\pi L}. \tag{S14}$$

Here $q_{iso} \approx (q_{11} - q_{22}) \approx q_{44} = 0.301 \times 10^{-12} \text{ Pa}^{-1}$ [S14], $n = 2.42$ is the diamond refractive index [S15], $L$ is the diamond thickness, $\lambda$ is the optical wavelength, and $|\delta|$ is magnitude of the optical retardance phase from $|\sin \delta|$.

Solving for $\sigma$ yields

$$|\sigma| \sim \frac{2}{3\pi} \frac{|\delta|\lambda}{Ln^3 q_{iso}}. \tag{S15}$$

A phase accumulation of $|\delta| = \pi/2$ gives $|\sigma| \sim 85$ MPa, which is the maximum stress we can measure without order ambiguity for an optical path length of 500 µm.

## Minimum detectable stress using birefringence imaging

The above method can yield a rough estimate for the minimum detectable stress using birefringence. A well-optimized system can detect down to $|\sin \delta| \approx 0.001$ [S10]. This gives an approximate [S11, S14] minimum stress detected using the isotropic photo-elastic constants and assuming a constant stress throughout the 500 µm thickness of diamond of $\sim 0.1$ MPa. More generally, we can write this as 50 MPa· µm for an arbitrary optical path length through the sample thickness. This value is typically limited by optical element quality, optical alignment, and employed calibration schemes. In our birefringence imager we can detect a minimum of $|\sin \delta| \approx 0.005$, giving a minimum-detectable stress of 0.5 MPa (250 MPa· µm). This value is limited by the employed optical elements, LED illumination power, and residual misalignment of the circular analyzer.

## Comparing birefringence and NV $M_{z,\kappa}$ measurements for Sample A

We used Sample A for a quantitative comparison of stress magnitudes generated using the NV $M_{z,\kappa}$ and optical birefringence methods. This sample contains regions which adequately fulfill the approximations needed for the rough isotropic model. It displays a relatively uniform spatial strain pattern in the $X$-$Y$ imaging plane. Furthermore, Sample A is polished on its top and bottom faces and one of its side faces. This allows us to partially image its birefringence through the sides of the diamond. These measurements, shown in Fig. S5, show a generally uniform depth distribution of strain, validating the use of the birefringence-to-stress approximate model above.

One of the current limitations of the birefringence method is order ambiguity in the measurement of $|\sin \delta|$. For high-strain regions, the accumulated phase difference through the sample thickness is greater than $\pi/2$, leading to ambiguity when calculating $\delta$ and stress from $|\sin \delta|$. As described above, the maximum stress magnitude is $\sim 85$ MPa for $L = 500$ µm and $\lambda = 532$ nm (Eq. S15). To unambiguously determine the stress magnitude from birefringence in samples with larger stress would require thinner (smaller $L$) samples or alternative methods [S16]. An example of this type of phase ambiguity can be seen for a sample region of interest in Sample A shown in Fig. S6 (the diagonal red stripe on the left side of the image) where $|\sin \delta|$ reaches its maximum value of 1 before decreasing in the middle of the stripe.

Figure S7 compares the birefringence map with the NV $\sigma_{XY}$ map for a petal-shaped strain defect. The birefringence data is plotted as $\sin^{-1}(|\sin \delta|)$ to make all images linear in the strain amplitude. Although both images show a strain feature with four lobes, the shapes are qualitatively dissimilar, with one rotated $45°$ with respect to the other. The lobe orientations in the $|\sin \delta|$ map are consistent with a dislocation bundle [S17]. In birefringence imaging and X-ray topography studies from Ref. [S18] dislocation bundles were found to diverge as they propagated through the diamond thickness. Such divergence may explain the difference in the spatial structure observed between the $|\sin \delta|$ map, which is integrated over the full 500 µm thickness, and the $\sigma_{XY}$ map, which shows stress in only the top 13 µm of the diamond (Fig. S7). In addition to the large petal feature, we also observe smaller petal defects in the

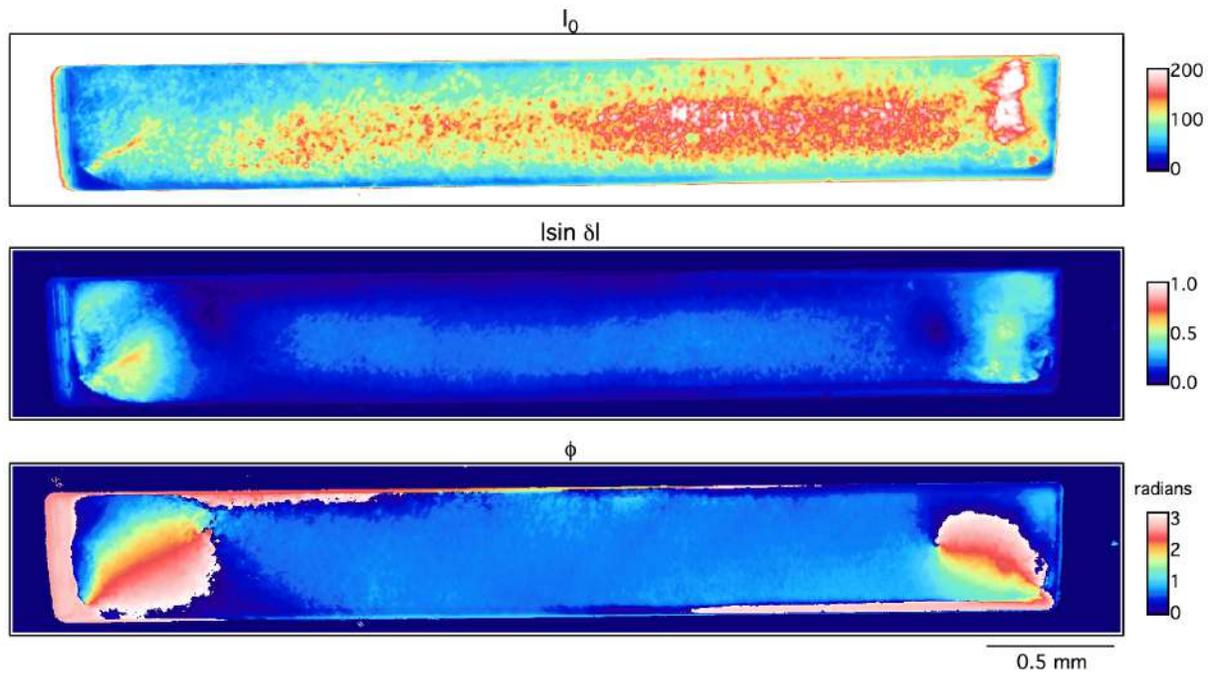

**Supplemental Figure** S5. Birefringence maps measured through a side face of Sample A. The speckle in the transmission image is due to one of the edges of the diamond in the optical path not being polished. This limits what can be quantitatively said about the distribution of strain, however, sources can be roughly localized. The stripe in the lower-left corner of Fig. 3 is on the left side of the image.

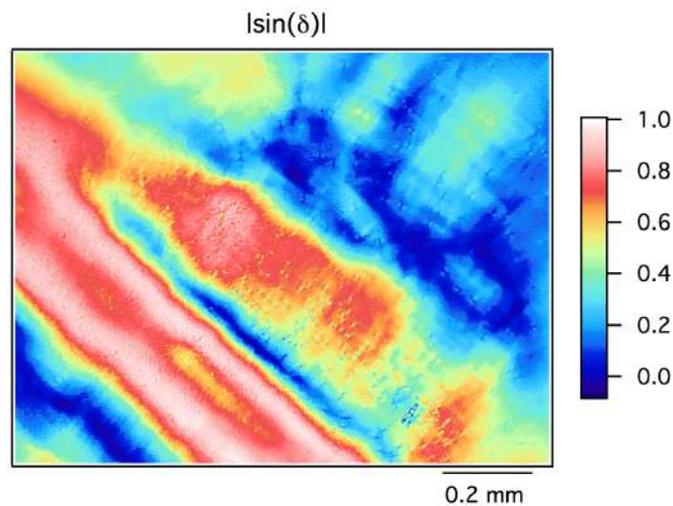

**Supplemental Figure** S6. A $|\sin\delta|$ birefringence map for a diagonal stripe strain defect in Sample A (see Fig. 4 in the main text). Despite the order ambiguity, the maximum stress measured using the birefringence and the $M_{z,\kappa}$ imaging techniques are consistent.

$\sigma_{XY}$ maps that are not present in the birefringence maps. These features likely originate at the interface of the 13 μm thick N-doped layer with the diamond substrate; the short vertical extent of these features put them below the birefringence sensitivity limit.

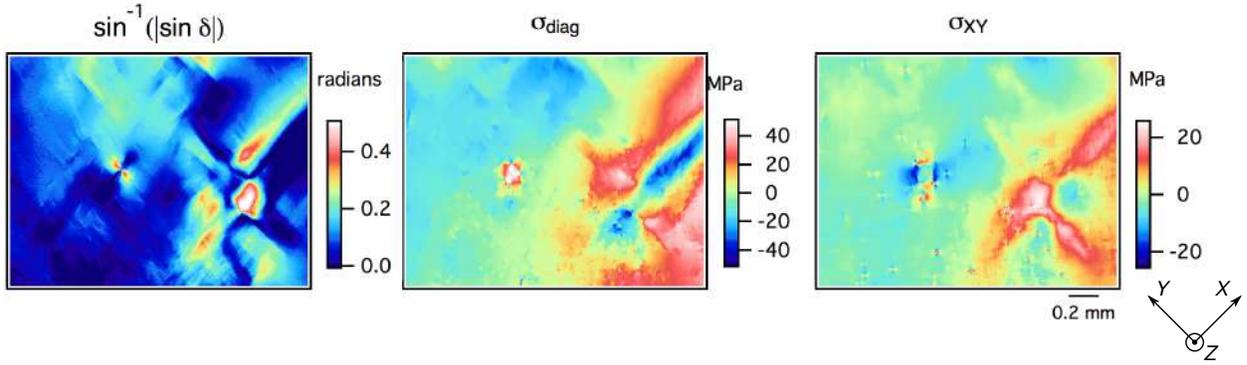

**Supplemental Figure** S7. A zoomed-in region on Sample A showing a large petal-shaped defect originating from a dislocation bundle, imaged with birefringence and NV $M_{z,\kappa}$ imaging. The NV measurement is only sensitive to the stress in a 13 μm layer, and shows some smaller petal-shaped defects not observed the birefringence measurement.

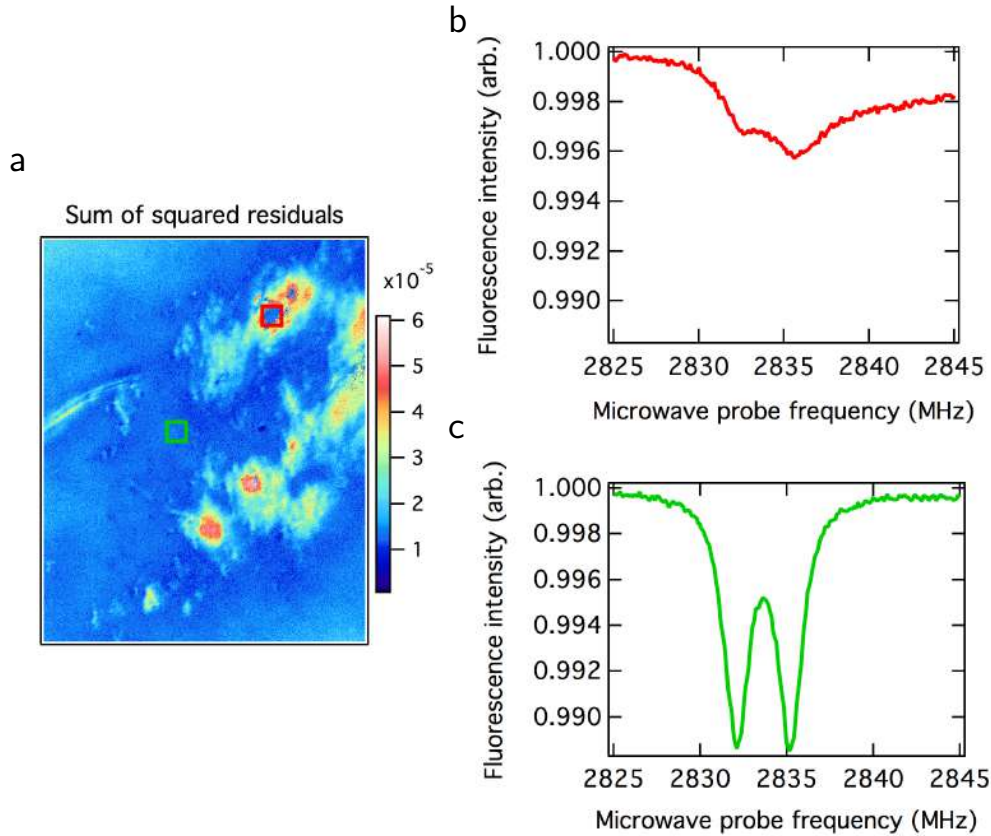

**Supplemental Figure** S8. (a) Sum of squared residuals for a region of Sample C. (b)-(c) ODMR lineshapes for the $m = 0$ to $-1$ transition for a high-strain (red) and low-strain (green) region. The locations of these regions within the image are represented in (a) with red and green boxes. In addition to causing lineshape broadening, the strain gradient in each pixel leads to worse fit residuals. Each NV resonance is split into two lines due to hyperfine interactions with the spin-1/2 $^{15}$N nucleus.

## STRAIN LIMITATIONS FOR NV MAGNETIC AND $M_{z,\kappa}$ IMAGING

### Influence of $M_{z,\kappa}$ gradients on NV resonance lineshapes

As shown in Fig. S8, $M_{z,\kappa}$ gradients can distort the single-pixel ODMR spectra, making lineshape fitting more difficult. For large gradients, the Lorentzian fit functions we use to extract the ODMR center frequencies may not describe the $M_{z,\kappa}$-broadened lineshapes accurately, potentially causing strain features to appear as systematic errors

in the magnetic images. The strain gradients within each pixel in Sample C lead to worse fit residuals (suggesting that a Lorentzian fit function is not the best choice), and also cause lineshape broadening. Ideally the single-pixel $M_{z,\kappa}$ inhomogeneity should be small compared to the other factors that determine the NV resonance linewidth (magnetic inhomogeneity, microwave field strength, and laser intensity), which is often ~1 MHz full width at half maximum.

### $M_{z,\kappa}$ inhomogeneity effects on magnetic vector measurements

To further emphasize the importance of mapping and minimizing strain in the diamond, Fig. S9 shows $M_{z,\kappa}$ maps for a lattice defect in Sample D, along with the corresponding vector magnetic images in the presence of a spatially-uniform bias magnetic field. Although the magnetic map should be homogeneous, the strain defect appears as a false magnetic feature. The false feature may be caused by covariance between fit parameters (fluorescence contrasts, resonance linewidths, and resonance frequencies) or by the $M_{z,\kappa}$ inhomogeneity distorting the ODMR lineshapes. This pathological example shows how the $M_{z,\kappa}$ imaging technique allows identification and rejection of magnetic features caused by strain, while also emphasizing the value of diamond samples with homogeneous stress/strain for NV magnetic imaging.

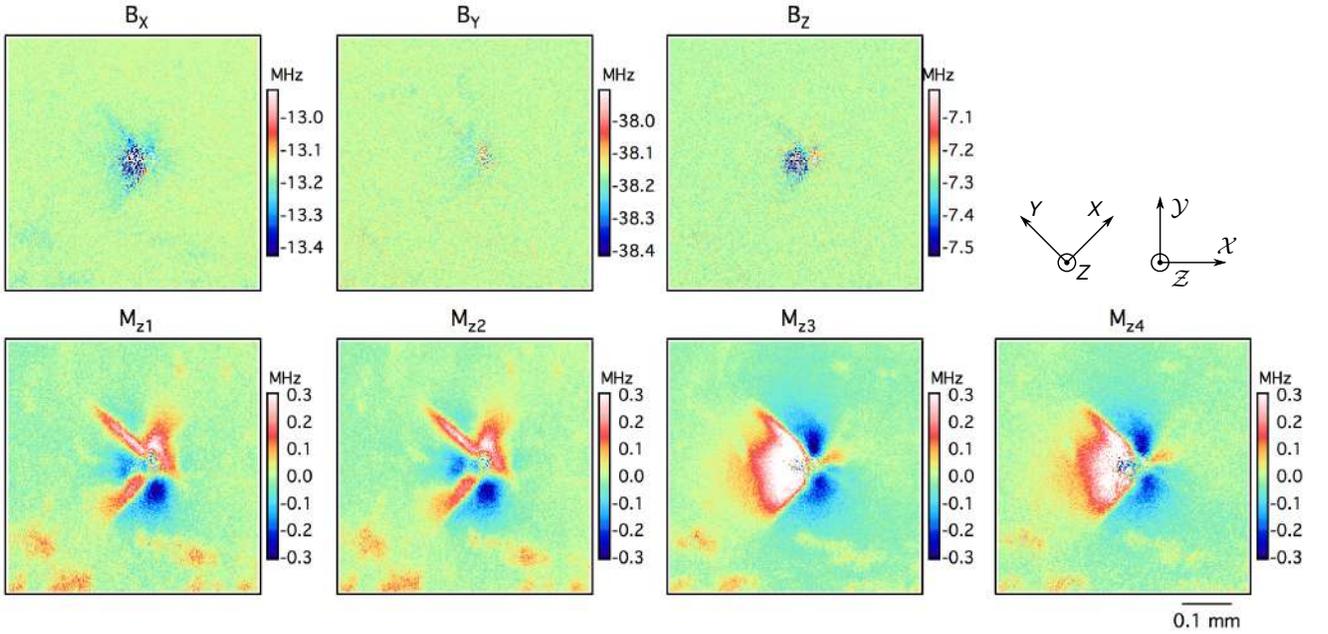

**Supplemental Figure** S9. Zoomed-in image of the $M_{z,\kappa}$ maps for a lattice defect feature in Sample D, together with the vector magnetic field maps from the same experiment in the presence of a spatially-uniform bias magnetic field. All color scales are in MHz (divide by 28.03 GHz/T for magnetic maps in tesla). The coordinates for $\{B_{\mathcal{X}}, B_{\mathcal{Y}}, B_{\mathcal{Z}}\}$ correspond to the diamond chip coordinate system $\{\mathcal{X}, \mathcal{Y}, \mathcal{Z}\}$, which is rotated by $45°$ from the $\{X, Y, Z\}$ unit cell coordinate system used for the stress maps. This example shows how a sufficiently large stress inhomogeneity can generate a false magnetic feature.

### Advantages of NV $M_{z,\kappa}$ imaging

Here we discuss technical advantages of NV $M_{z,\kappa}$ imaging over birefringence imaging.

- An NV $M_{z,\kappa}$ imager measures the stress within the microscope depth-of-field, localizing the depth to within few micrometers in the optical diffraction limit or up to a few nanometers in a shallow nitrogen implant or delta-doped layer [S19]. A birefringence imager integrates optical retardance through the entire diamond thickness (typically hundreds of micrometers), making depth information difficult to assess.

- Since optical retardance increases with diamond thickness, birefringence imaging can be difficult for thin diamonds (30 μm thickness). Similarly, order ambiguity complicates analysis of birefringence measurements with thick diamond samples when $|\delta| > \pi/2$.

- Extracting $\Delta n$ and stress from a birefringence image requires knowledge of $n$ and $L$. Other polarization effects like circular birefringence can lead to incorrect conclusions about the magnitude of the linear birefringence [S20].

- While a birefringence measurement is useful for qualitative comparisons and phenomenology, an NV $M_{z,\kappa}$ measurement is more quantitative and easier to interpret. Interpreting $\Delta n$ information into an absolute stress or strain is model dependent [S11].

- With a birefringence imager, the extracted optical retardance depends on the illumination wavelength(s), polarizer orientations, and birefringence in other optical components or glass. An NV $M_{z,\kappa}$ imager avoids these complications, such that measured stress inhomogeneity maps should be independent of the fluorescence microscope employed.

- Detecting whether a diamond is uniformly compressed or dilated with a birefringence measurement is challenging. An NV $M_{z,\kappa}$ measurement would see compression and dilation as a uniform shift of all four $M_{z,\kappa}$ parameters [S21]. Similarly, while $M_{z,\kappa}$ images yield all shear stress terms, birefringence imaging is only sensitive to $\sigma_{XY}$ shear terms (in the plane perpendicular to the optical axis) for a given measurement [S22].

- NV $M_{z,\kappa}$ imaging requires optical access from one polished side of the diamond. Birefringence imaging requires an uninterrupted optical path through the diamond, which may be impossible if the diamond is mounted on an opaque substrate or if the back surface is too rough.

### Advantages of birefringence imaging

In spite of the above drawbacks, birefringence imaging may be preferable under certain circumstances.

- Birefringence imaging is a faster experiment; it requires at minimum a single camera image (or a few images if using multiple waveplate orientations). An NV $M_{z,\kappa}$ measurement needs a few hundred camera images (roughly 50 microwave probe frequencies per NV frequency). In addition, obtaining a sufficient signal-to-noise ratio to fit the NV ODMR lineshape in each pixel usually requires averaging. This makes birefringence imaging useful for quickly characterizing diamond substrates.

- A birefringence measurement has one main parameter to optimize before measuring: namely, the $\lambda/4$ waveplate and linear polarizer should be maintained at $45°$ with respect to one another. For an NV experiment, the laser polarization, microwave field, and bias magnetic field must all be optimized.

- Just as strain inhomogeneity can complicate an NV magnetic microscopy experiment, magnetic field inhomogeneity can complicate an NV $M_{z,\kappa}$ experiment. A birefringence experiment is insensitive to magnetic fields.

- NV $M_{z,\kappa}$ imaging is difficult in diamonds with low NV density, broad ODMR linewidths, or weak fluorescence contrast. One example of low-NV diamonds are the high-purity diamond samples used as substrates for growth of nitrogen-enriched layers. It is imperative that these seed diamonds contain low strain inhomogeneity, as strain defects tend to propagate into the nitrogen-enriched layer. Birefringence imaging is a more fitting choice to characterize strain in such high-purity diamonds.

- In diamonds with a thick NV layer (compared to the microscope depth-of-field), fluorescence light from out-of-focus NVs can enter the microscope, complicating the ODMR $M_{z,\kappa}$ interpretation and potentially blurring the $M_{z,\kappa}$ map. This phenomenon is visible in Sample C, where the stress maps are blurred by the 40 μm NV layer when compared to the birefringence measurements.

### NV $M_{z,\kappa}$ AND STRESS MAPS FOR ADDITIONAL SAMPLES

Figures S10-S19 include the complete NV $M_{z,\kappa}$ and stress maps for all diamond samples studied in this work (Samples A through J). Table S1 lists the properties of these diamonds. Here we comment on the strain features seen in each sample.

- Sample B has mm-sized X-shaped strain features present in the $\sigma_{\text{diag}}$ and $\sigma_{XY}$ maps (also seen in Fig. S2).

- Sample C has a 40 μm NV layer, which can degrade the NV stress mapping spatial resolution if the camera collects fluorescence from the entire NV layer. Comparing Fig. S4 and Fig. S12, if the NV layer is thicker than the typical strain feature size, the birefringence maps may yield a sharper strain image. Sample C is unique in that it has a comparable inhomogeneity in all of the stress components throughout the entire 2D layer, which could be caused by mechanical polishing and etching.

- Sample D is a nitrogen ion ($^{14}N^+$) implant diamond. This sample was cut from a larger diamond chip along the left and bottom edges, which exhibit less broad-scale plastic deformation than the original (top and right) edges. This sample also has some finer-scale strain inhomogeneity along the bottom edge.

- Sample E has strain inhomogeneity caused by petal-shaped defects, broad-scale plastic deformation, and a relatively homogeneous region in the middle of the sample. A larger sample was laser-cut into quarters, one piece of which is Sample E. This piece has less stress near the middle and at the cut edges (right and bottom) than at the uncut edges (left and top).

- Sample F has undergone a fracture on the right side of the image (a broken corner, see Fig. 1c in Ref. [S2]). Having strain and broad-scale plastic deformation associated with sharp corners, edges, and fractures is to be expected. The pixels in the top right have a $M_{z,\kappa}$ gradient large enough to prevent lineshape fitting using our standard fit function, and are set to zero in this image. The $M_{z,\kappa}$ values are more homogeneous toward the diamond chip interior. The $\sigma_{XZ}$ and $\sigma_{YZ}$ stress contributions are smaller compared to the $\sigma_{diag}$ and $\sigma_{XY}$ contributions, but still include some fine-scale features.

- Sample G has square etch pits [S18] and irregularly-shaped indentations on the diamond surface. These surface imperfections can come from preferential etching of lattice dislocation tracks or from damage due to mechanical polishing, and can therefore be considered as symptoms or indicators of diamond strain. The irregularly-shaped surface indentations are correlated with diamond stress (mostly in $\sigma_{diag}$ and $\sigma_{XY}$).

- Sample H has diagonal streaks in the strain maps. Note that due to suboptimal experimental conditions the $M_{z,\kappa}$ and stress maps include some ripple artefacts, which are correlated with the inhomogeneous laser illumination. Fortunately, the ripple artefacts do not affect the stress maps significantly.

- Sample I has several mm-sized X-shaped strain features, similar to those seen in other samples.

- Sample J has $M_{z,\kappa}$ and stress inhomogeneity from diagonal stripes and from the top and left edges.

---

* Current address: Sandia National Laboratories, Albuquerque, NM 87123, USA

[S1] D. R. Glenn, R. R. Fu, P. Kehayias, D. Le Sage, E. A. Lima, B. P. Weiss, and R. L. Walsworth, Geochemistry, Geophysics, Geosystems 18, 3254 (2017).

[S2] E. Bauch, C. A. Hart, J. M. Schloss, M. J. Turner, J. F. Barry, P. Kehayias, S. Singh, and R. L. Walsworth, Phys. Rev. X 8, 031025 (2018).

[S3] H. C. Davis, P. Ramesh, A. Bhatnagar, A. Lee-Gosselin, J. F. Barry, D. R. Glenn, R. L. Walsworth, and M. G. Shapiro, Nature Communications 9, 131 (2018).

[S4] M. P. Backlund, P. Kehayias, and R. L. Walsworth, Phys. Rev. Applied 8, 054003 (2017).

[S5] M. S. J. Barson, P. Peddibhotla, P. Ovartchaiyapong, K. Ganesan, R. L. Taylor, M. Gebert, Z. Mielens, B. Koslowski, D. A. Simpson, L. P. McGuinness, J. McCallum, S. Prawer, S. Onoda, T. Ohshima, A. C. Bleszynski Jayich, F. Jelezko, N. B. Manson, and M. W. Doherty, Nano Letters 17, 1496 (2017).

[S6] A. Barfuss, M. Kasperczyk, J. Kölbl, and P. Maletinsky, Phys. Rev. B 99, 174102 (2019).

[S7] P. Udvarhelyi, V. O. Shkolnikov, A. Gali, G. Burkard, and A. Pályi, Phys. Rev. B 98, 075201 (2018).

[S8] M. W. Doherty, F. Dolde, H. Fedder, F. Jelezko, J. Wrachtrup, N. B. Manson, and L. C. L. Hollenberg, Phys. Rev. B 85, 205203 (2012).

[S9] A. M. Glazer, J. G. Lewis, and W. Kaminsky, Proceedings of the Royal Society of London A: Mathematical, Physical and Engineering Sciences 452, 2751 (1996).

[S10] W. Kaminsky, E. Gunn, R. Sours, and B. Kahr, Journal of Microscopy 228, 153 (2007).

[S11] D. Howell, European Journal of Mineralogy 24, 575 (2012).

[S12] D. Howell, I. G. Wood, F. Nestola, P. Nimis, and L. Nasdala, European Journal of Mineralogy 24, 563 (2012).

[S13] D. Howell, I. G. Wood, D. P. Dobson, A. P. Jones, L. Nasdala, and J. W. Harris, Contributions to Mineralogy and Petrology 160, 705 (2010).

[S14] D. Howell, I. G. Wood, D. P. Dobson, A. P. Jones, L. Nasdala, and J. W. Harris, Contributions to Mineralogy and Petrology 160, 705 (2010).

[S15] A. M. Zaitsev, Optical Properties of Diamond: A Data Handbook (Springer, 2001).

[S16] Geday, Kaminsky, Lewis, and Glazer, Journal of Microscopy **198**, 1 (2000).

[S17] H. Pinto and R. Jones, Journal of Physics: Condensed Matter **21**, 364220 (2009).

[S18] I. Friel, S. Clewes, H. Dhillon, N. Perkins, D. Twitchen, and G. Scarsbrook, Diamond and Related Materials **18**, 808 (2009).

[S19] K. Ohno, F. J. Heremans, L. C. Bassett, B. A. Myers, D. M. Toyli, A. C. B. Jayich, C. J. Palmstrom, and D. D. Awschalom, Applied Physics Letters **101**, 082413 (2012).

[S20] H. Jasbeer, R. J. Williams, O. Kitzler, A. McKay, S. Sarang, J. Lin, and R. P. Mildren, J. Opt. Soc. Am. B **33**, B56 (2016).

[S21] M. W. Doherty, V. V. Struzhkin, D. A. Simpson, L. P. McGuinness, Y. Meng, A. Stacey, T. J. Karle, R. J. Hemley, N. B. Manson, L. C. L. Hollenberg, and S. Prawer, Phys. Rev. Lett. **112**, 047601 (2014).

[S22] H. Pinto, R. Jones, J. P. Goss, and P. R. Briddon, J. Phys.: Conf. Ser. **281**, 012023 (2011).

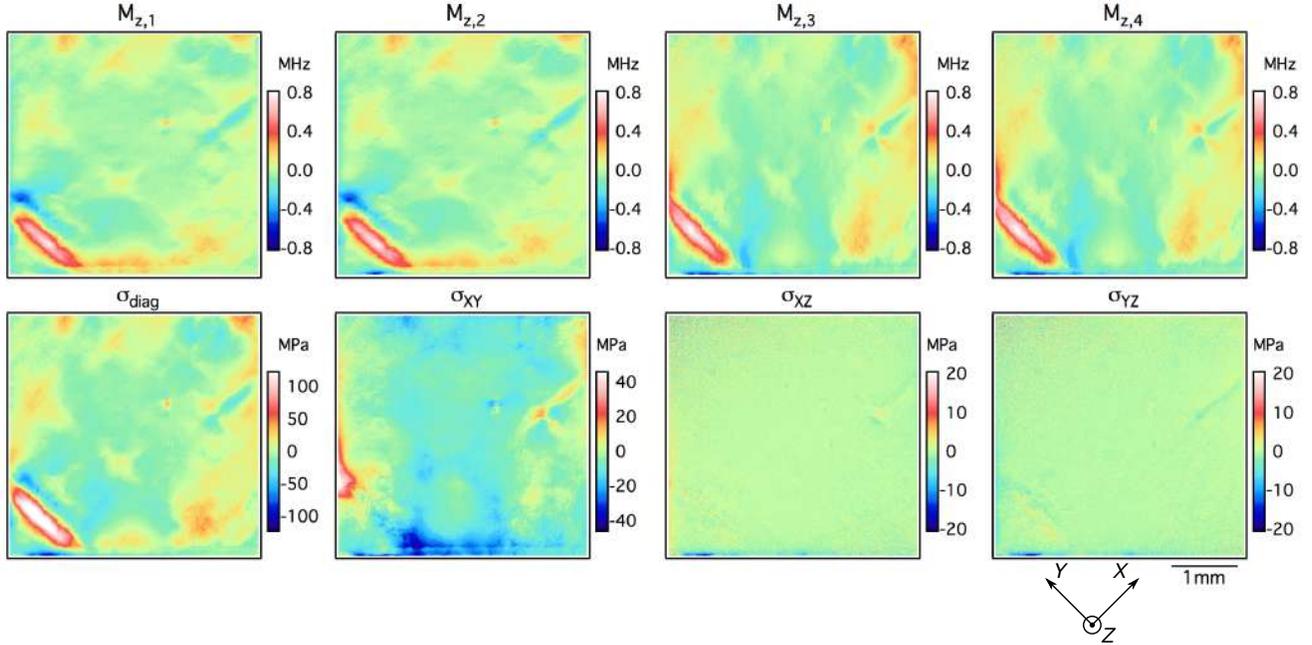

**Supplemental Figure** S10. NV $M_{z,\kappa}$ and $\{\sigma_{diag}, \sigma_{XY}, \sigma_{XZ}, \sigma_{YZ}\}$ maps for diamond Sample A. (Identical to Fig. 3 in the main text, but included here for completeness.)

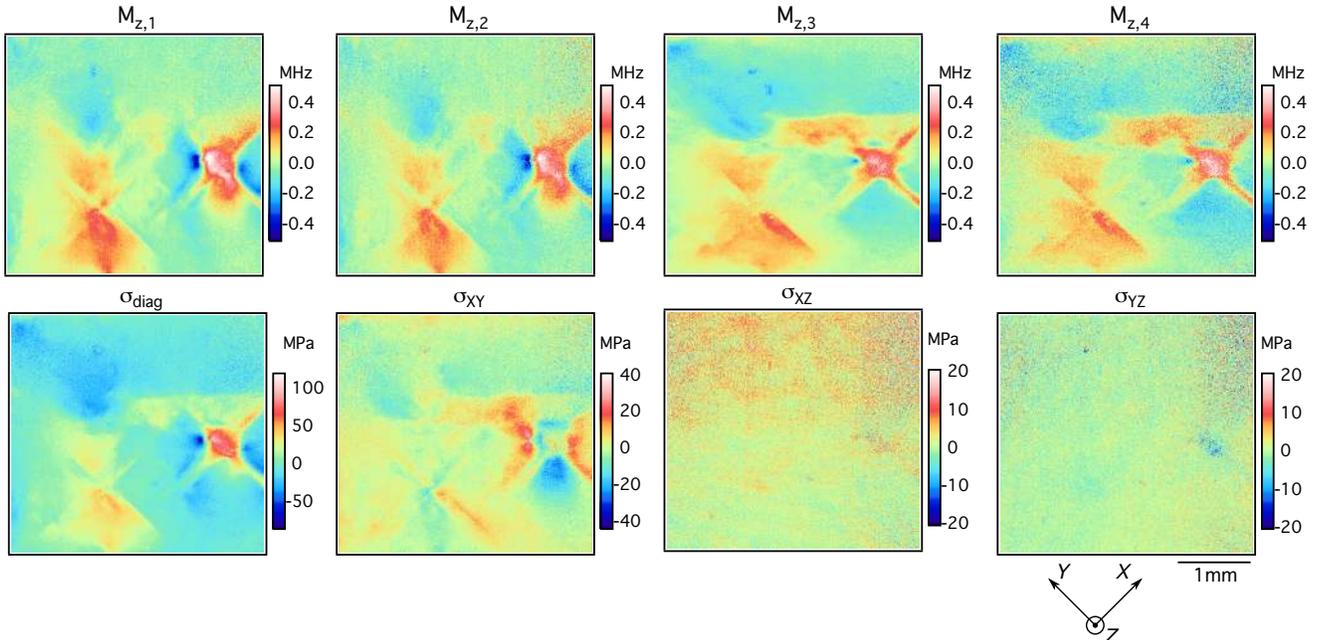

**Supplemental Figure** S11. NV $M_{z,\kappa}$ and $\{\sigma_{diag}, \sigma_{XY}, \sigma_{XZ}, \sigma_{YZ}\}$ maps for diamond Sample B.

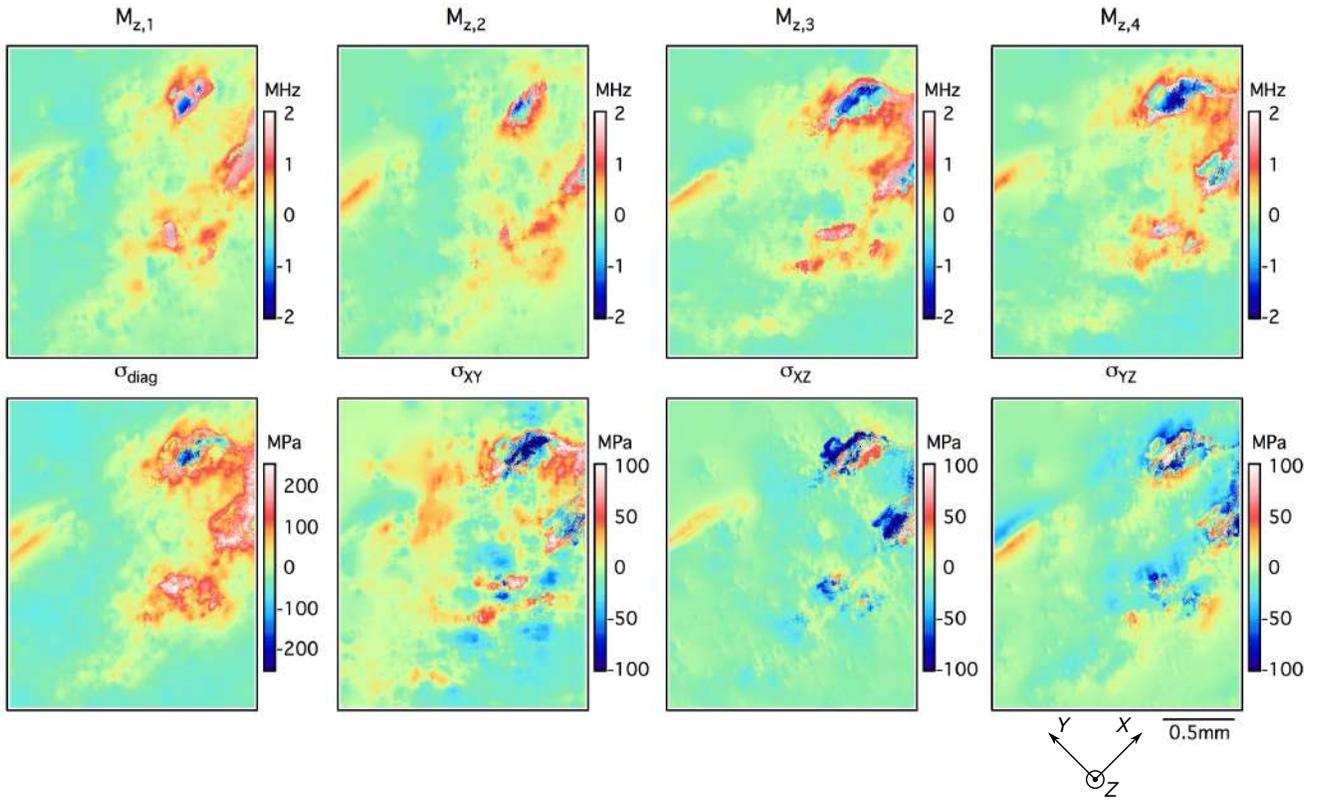

**Supplemental Figure** S12. NV $M_{z,\kappa}$ and $\{\sigma_{diag}, \sigma_{XY}, \sigma_{XZ}, \sigma_{YZ}\}$ maps for diamond Sample C.

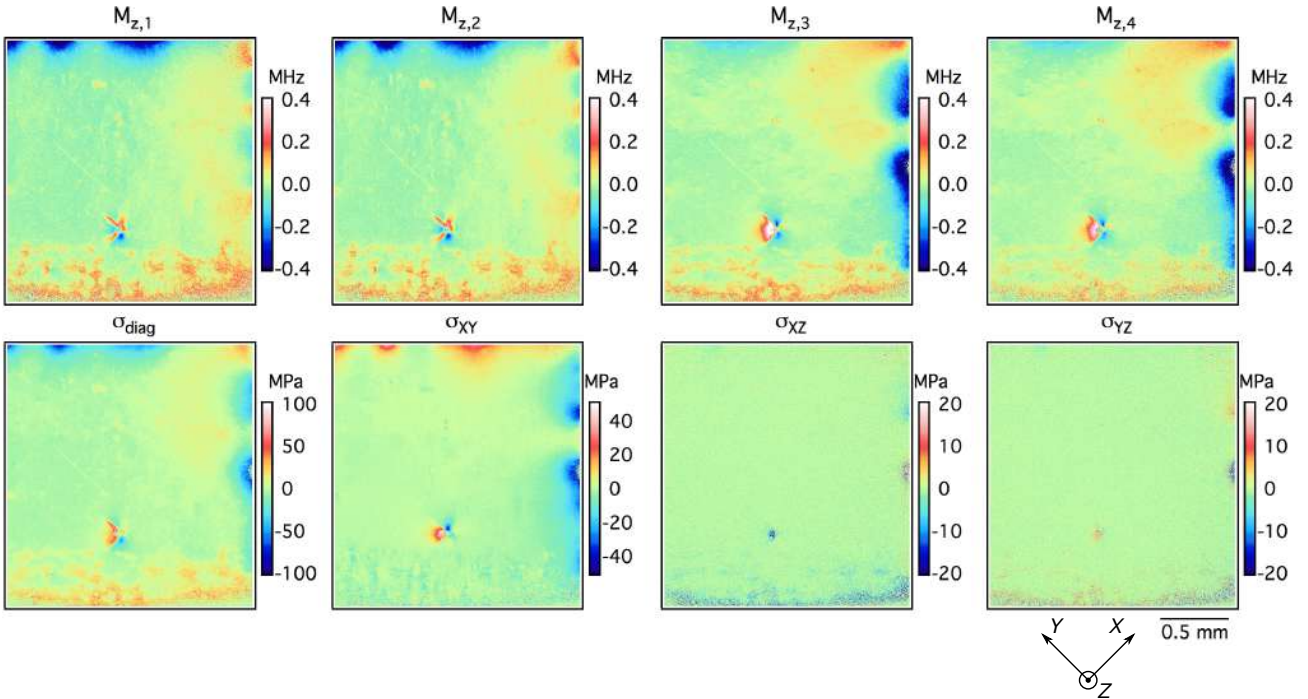

**Supplemental Figure** S13. NV $M_{z,\kappa}$ and $\{\sigma_{diag}, \sigma_{XY}, \sigma_{XZ}, \sigma_{YZ}\}$ maps for diamond Sample D.

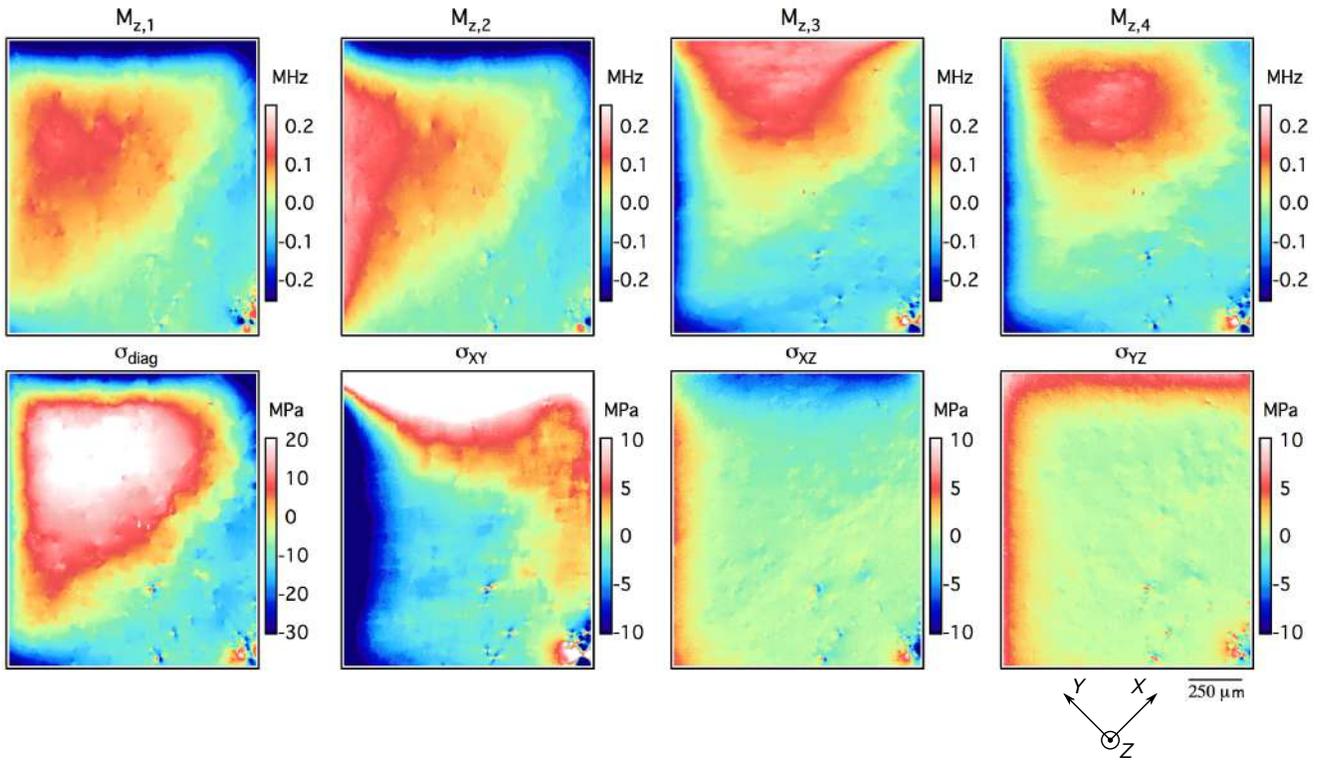

**Supplemental Figure** S14. NV $M_{z,\kappa}$ and $\{\sigma_{diag}, \sigma_{XY}, \sigma_{XZ}, \sigma_{YZ}\}$ maps for diamond Sample E.

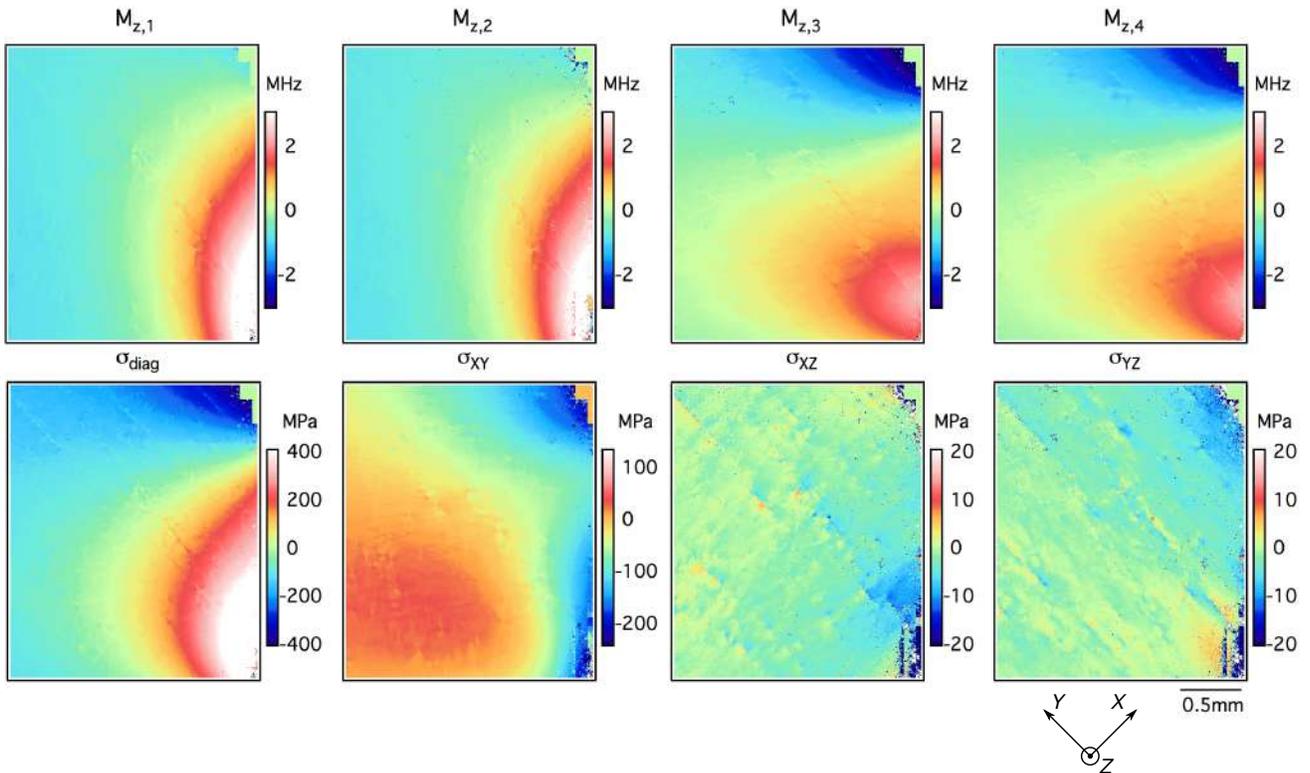

**Supplemental Figure** S15. NV $M_{z,\kappa}$ and $\{\sigma_{diag}, \sigma_{XY}, \sigma_{XZ}, \sigma_{YZ}\}$ maps for diamond Sample F.

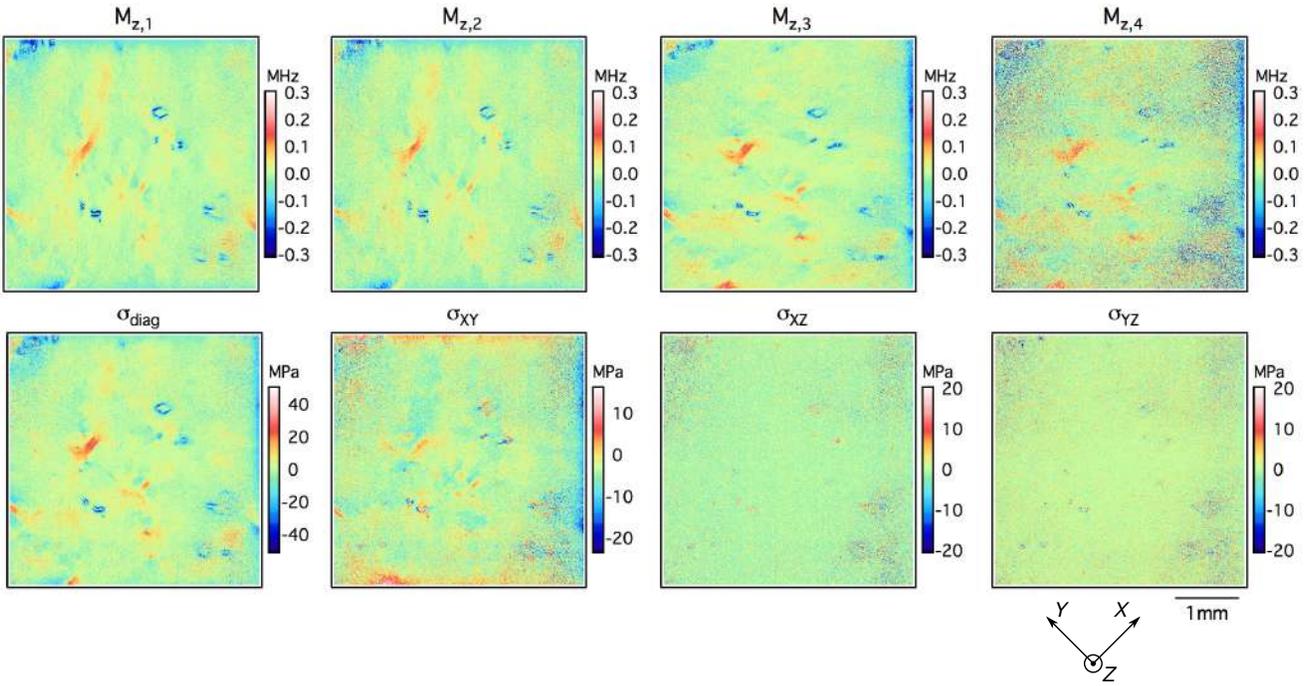

**Supplemental Figure** S16. NV $M_{z,\kappa}$ and $\{\sigma_{diag}, \sigma_{XY}, \sigma_{XZ}, \sigma_{YZ}\}$ maps for diamond Sample G.

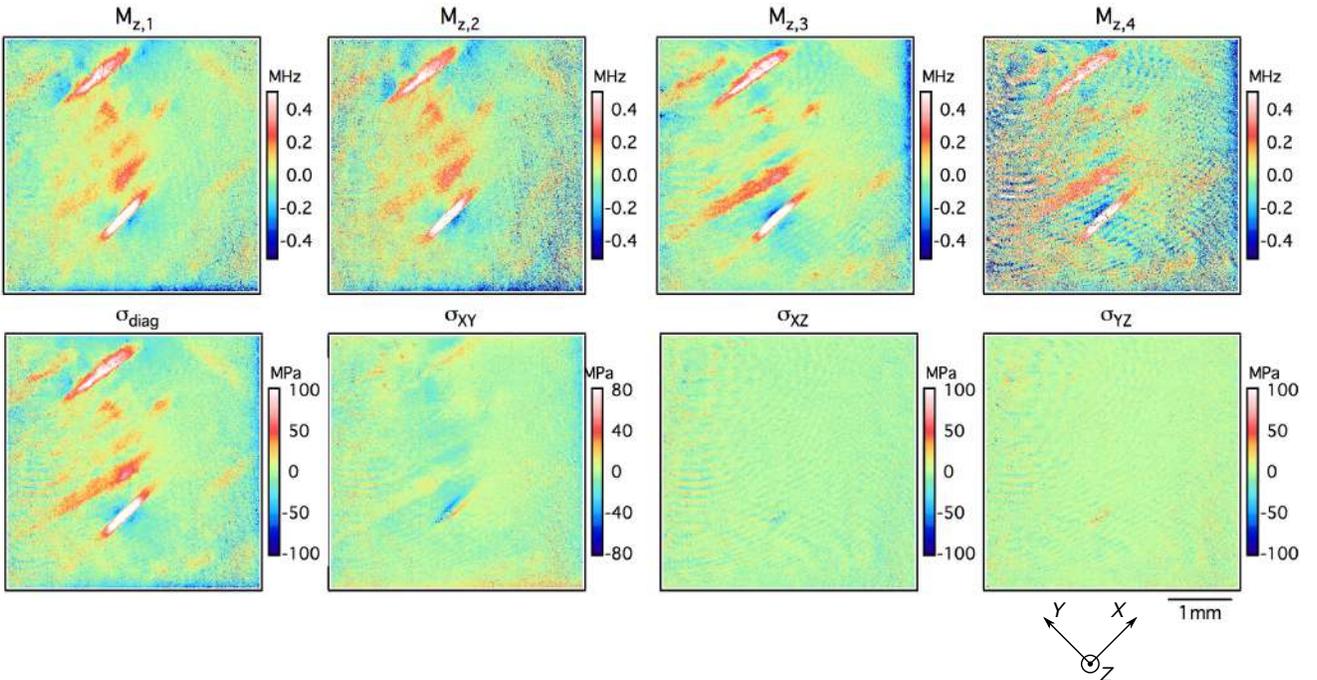

**Supplemental Figure** S17. NV $M_{z,\kappa}$ and $\{\sigma_{diag}, \sigma_{XY}, \sigma_{XZ}, \sigma_{YZ}\}$ maps for diamond Sample H.

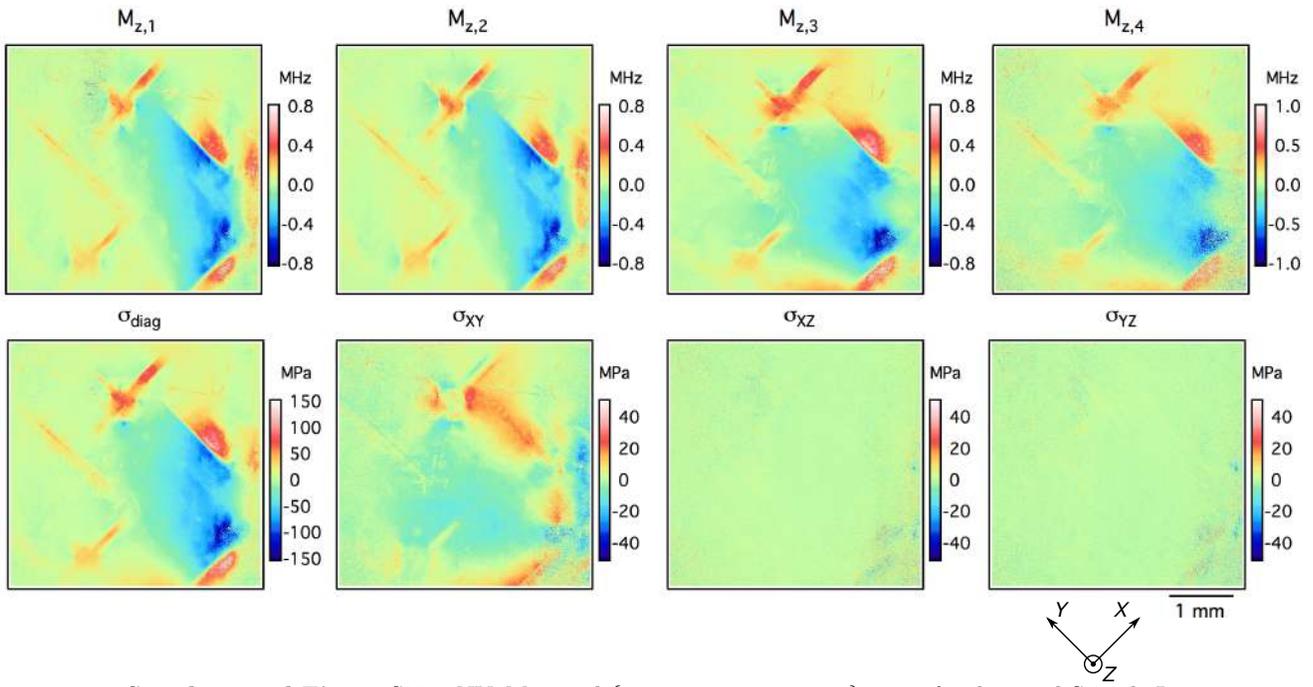

**Supplemental Figure** S18.  NV $M_{z,\kappa}$ and $\{\sigma_{diag}, \sigma_{XY}, \sigma_{XZ}, \sigma_{YZ}\}$ maps for diamond Sample I.

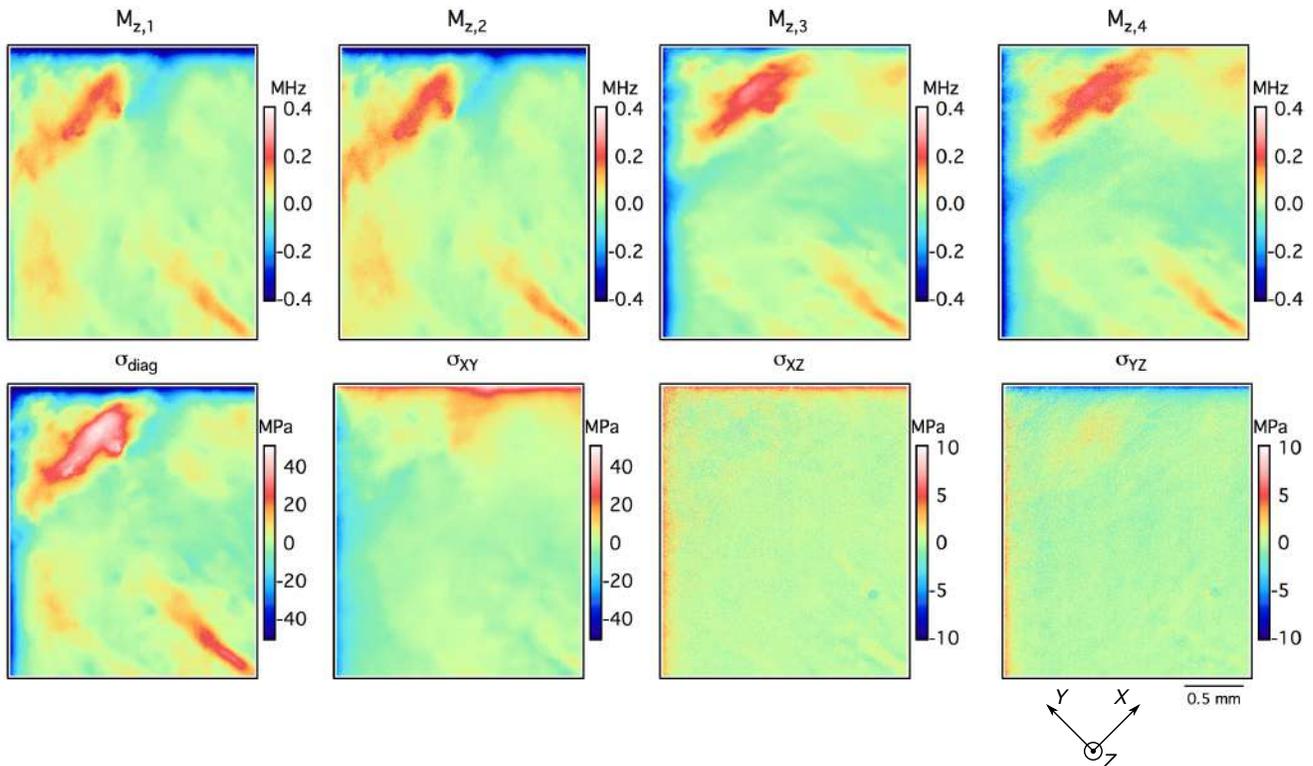

**Supplemental Figure** S19.  NV $M_{z,\kappa}$ and $\{\sigma_{diag}, \sigma_{XY}, \sigma_{XZ}, \sigma_{YZ}\}$ maps for diamond Sample J.



\end{document}